\documentclass[lettersize,onecolumn]{IEEEtran}
\usepackage{cite}
\usepackage{amsmath,amssymb,amsfonts}
\usepackage{algorithmic}
\usepackage{graphicx}
\usepackage{textcomp}
\usepackage{xcolor}
\usepackage{authblk}

\newtheorem{theorem}{\textbf{Theorem}}

\newtheorem{lemma}{\textbf{Lemma}}
\newtheorem{remark}{Remark}
\newtheorem{proposition}{\textbf{Proposition}}

\newcommand{\tr}{\text{Tr}}
\newcommand{\cov}{\text{Cov}}

\def\BibTeX{{\rm B\kern-.05em{\sc i\kern-.025em b}\kern-.08em
    T\kern-.1667em\lower.7ex\hbox{E}\kern-.125emX}}

\makeatletter
\renewcommand{\section}{\@startsection{section}{1}{0mm}
	{-\baselineskip}{0.5\baselineskip}{\bf\leftline}}
\makeatother

\begin{document}
\title{ Implicit Communication in Linear Quadratic Gaussian Control Systems }

\author{Gongpu Chen,~\IEEEmembership{Member,~IEEE} and Deniz Gunduz,~\IEEEmembership{Fellow,~IEEE}
	% <-this % stops a space
	\thanks{The authors are with the Department of Electrical and Electronic Engineering, Imperial College London, London SW7 2AZ, UK (e-mails:\{gongpu.chen, d.gunduz\}@imperial.ac.uk).}% <-this % stops a space
	%\thanks{Manuscript received October 26, 2023; revised December 8, 2023.}
}

\maketitle

\begin{abstract}
	This paper studies implicit communication in linear quadratic Gaussian control systems. We show that the control system itself can serve as an implicit communication channel, enabling the controller to transmit messages through its inputs to a receiver that observes the system state. This communication is considered implicit because (i) no explicit communication channels are needed; and (ii) information is transmitted while simultaneously fulfilling the controller’s primary objective—maintaining the control cost within a specified level. As a result, there exists an inherent trade-off between control and communication performance. This trade-off is formalized through the notion of implicit channel capacity, which characterizes the supremum reliable communication rate subject to a constraint on control performance.
	We investigate the implicit channel capacity in two settings. When both the controller and the receiver have noiseless observations of the system state, the channel capacity admits a closed-form expression. When both the controller and the receiver have noisy observations, we establish a lower bound on the implicit channel capacity. Surprisingly, in the noiseless observation case, the capacity-achieving input policy adheres to a separation principle, allowing the control and channel coding tasks to be addressed independently, without loss of optimality. While this separation principle no longer holds in the noisy observation setting, we show that linear Gaussian input policies still decouple the channel coding problem from control, and can thus greatly simplify the practical implementation of implicit communication.
    %Moreover, under this capacity-achieving input policy, the implicit channel can be equivalently translated into a Gaussian MIMO channel, enabling the use of existing channel codes to achieve implicit communication.
	
\end{abstract}

\begin{IEEEkeywords}
	Implicit communication, linear quadratic Gaussian control, channel capacity, separation principle
\end{IEEEkeywords}

\section{Introduction}  \label{sec:intro}
\IEEEPARstart{I}{mplicit} communication is widespread in nature and human society, with examples including coordination and signaling in starling murmurations \cite{ballerini2008interaction}, fish schooling \cite{katz2011inferring}, human body language \cite{Trenholm:20,beattie2016rethinking}, and more \cite{Waters:ARCDB:05, Thienen:BES:14, locusts2025}. These phenomena demonstrate that information can be conveyed through observable actions and behaviors, without relying on explicit communication channels.
This concept is also appealing in artificial systems, particularly in scenarios where explicit communication is unavailable or where implicit communication can serve as an effective complement.
For instance, in autonomous driving \cite{fisac2019hierarchical,schwarting2019social}, a vehicle may need to express its intent—such as parking in a specific spot, changing lanes, or dropping off a passenger—to surrounding vehicles to enable safe and coordinated interactions. In human–robot interaction \cite{knepper2017implicit,breazeal2005effects,dragan2013legibility}, a robot may need to signal its intent to facilitate smoother cooperation. In swarm robotics \cite{connor2020current,swarm2021,d2005making}, robots often need to exchange information to coordinate their actions, much like birds in flocks or insects in swarms. Implicit communication through actions and behaviors is a promising approach in all of these scenarios, as it mirrors the strategies naturally employed by humans and animals.

From an information-theoretic perspective, all communication—whether explicit or implicit—requires a channel: an input–output system that allows the receiver to infer, from the observed outputs, the message that is encoded into the inputs by the transmitter. In this sense, existing studies often lack a rigorous definition of implicit communication \cite{d2005making,liang2019implicit,tian2020learning}. What constitutes the channel in implicit communication? How is information transmitted through this channel? To address these questions, we first propose the following definitions: 
An \textit{explicit communication channel} is an input-output system dedicated solely to communication between a transmitter and a receiver.
In contrast, an \textit{implicit communication channel} is a system in which the inputs and outputs serve a primary functional role (e.g., control or decision-making) and simultaneously transmit information that can be decoded by a receiver observing the system outputs.
In such cases, communication emerges as a ``byproduct'' of the system’s operation through slight but purposeful modifications of the inputs and outputs—without the need for an external signaling mechanism.

This paper investigates implicit communication in linear quadratic Gaussian (LQG) control systems, where a linear system driven by Gaussian noise evolves under the influence of the controller’s inputs.
While the system is primarily designed to perform a control task, we show that it can also function as an implicit communication channel, allowing the controller to transmit messages to a receiver that can observe the system state, without relying on explicit channels. Specifically, the controller can encode the message into the control inputs, and the receiver then decodes the message from its observations of the system state---either noiseless or noisy. However, there is no free lunch—the implicit communication comes at the cost of some degradation in control performance. This is natural, as encoding messages into control inputs typically requires the controller to deviate from the optimal control policy. As a result, there exists a fundamental trade-off between control and communication performance.

We define the implicit communication channel within this framework using standard information-theoretic principles and formulate the implicit communication problem as a co-design of channel coding and control. The trade-off between control and communication is then characterized by the capacity of the implicit channel, subject to a constraint on the control performance. Our main contributions are theoretical analysis of the implicit channel capacity in two settings, as summarized below: 
\begin{itemize}
	\item [1.] We begin by analyzing the case where both the controller and the receiver have noiseless observations of the system state, and we derive a closed-form expression for the implicit channel capacity. The expression coincides with that of the memoryless Gaussian MIMO channel capacity. We show that, under the capacity-achieving input policy, the implicit channel is equivalent to a memoryless Gaussian MIMO channel, where the noise is exactly the process noise, and the control constraint translates into a power constraint on the channel input.  Moreover, in the case of scalar systems, the implicit channel capacity admits an explicit expression as a function of the optimal control cost and the tolerable control loss, clearly revealing the trade-off between control and communication.
	
	\item [2.] We then extend our analysis to the more general setting where the controller and the receiver observe the system state through different noise processes. In this case, the implicit channel becomes a memory channel with states and operates under noisy feedback, making capacity characterization particularly challenging. We derive a lower bound on the implicit channel capacity, which is achievable using a stationary linear Gaussian input policy. Moreover, we show that under such a policy, the implicit channel can be equivalently transformed into a Gaussian channel with memory—but without state information or feedback—making it more amenable to analysis and practical coding schemes.

	%that adhere to a separation principle, making it of significant practical relevance.
\end{itemize}

A surprising finding is the emergence of a separation principle in the structure of the optimal input policy when the controller and the receiver have noiseless observations—the setting in which we can fully characterize the channel capacity. In this case, we show that the capacity-achieving input policy is given by the sum of the optimal LQG control input and a Gaussian random variable. Since the optimal control input is fixed by the LQG control objective, implicit communication is achieved by encoding the message into the Gaussian component, which we refer to as the communication signal. In other words, in an LQG control system, implicit communication can be achieved by simply adding a communication signal to the optimal control input.
This separation principle significantly simplifies the problem, as it implies that the control and channel coding tasks can be addressed independently, without loss of optimality. 

Although this separation principle no longer holds in observations are noisy, the stationary linear Gaussian structure remains a desirable property for the input policy. With this structure, once the feedback gain is fixed, the implicit channel reduces to a Gaussian channel with a power constraint on the communication signal, allowing the coding problem to be addressed independently of the control design.
The lower bound we derive corresponds exactly to the maximum rate achievable by the input policies that adhere to this linear Gaussian structure, making the bound of significant practical importance.

The rest of the paper is organized as follows. Section \ref{sec:related} reviews the related work. Section \ref{sec:system} introduces preliminaries on LQG control and formulates the implicit communication problem in the noiseless setting. Section \ref{sec:perfect} presents the implicit channel capacity and the achievability proof for the noiseless case. Section \ref{sec:no} generalizes the analysis to the noisy observation setting and establishes a lower bound on the channel capacity. Section \ref{sec: proofs} presents the proofs of the main theorems.  Finally, Section \ref{sec:con} concludes this paper and discusses possible directions of future work.

\subsection{Notations}
For any sequence $\{x_t: t\ge 1 \}$, we will use $x^j_{i}$ to denote the subsequence $\{x_i, x_{i+1},\cdots,x_j \}$. Specially, $x_1^j$ will be simply written as $x^j$. Given any positive integer $n$, let $[n]=\{1,2,\cdots,n \}$ be the set of integers between $1$ and $n$. For a square matrix $A$, $\tr(A)$ and $\det(A)$ denote its trace and determinant, respectively. Positive definite (PD) and positive semi-definite (PSD) matrices are denoted by $A\succ 0$ and $A\succeq 0$, respectively. The pseudo inverse of a matrix $L$ is denoted by $L^\dagger$, whenever it exists. The identity matrix is denoted by $I$. Finally, following standard notation, $\mathcal{I}(\cdot;\cdot)$ denotes mutual information, $H(\cdot)$ and $h(\cdot)$ denote entropy and differential entropy, respectively.

\section{Related Work} \label{sec:related}
Implicit communication has been studied across various fields \cite{stegmann2013animal,albuquerque2022dogs,patterson2023four,denham2013beyond}, with the most relevant to our work being research in control-related domains such as multi-agent systems \cite{Shaw2022,wanglearning,sokota2022communicating} and human–robot interaction \cite{8967120,dey2017pedestrian,li2017implicit}, where the transmitter is an agent engaged in a control task. Most of these studies are experimental in nature and often rely on machine learning methods, lacking a clear formulation of communication and a fundamental understanding of the complex relationship between control and communication.

One exception is the work in \cite{Act2comm}, which models the environment of a Markov decision process (MDP) as a finite-state channel and analyzes the trade-off between the MDP reward and channel coding rate.  The channel capacity under a reward constraint is characterized as the solution to a convex optimization problem, where the objective function is the conditional mutual information $\mathcal{I}(A;S'|S)$, with $A$ denoting the action, and $S$ and $ S'$ representing the current and the next states, respectively. The limitation of \cite{Act2comm} is that it focuses on MDPs with finite state and action spaces, and only consider the setting with noiseless observations. In contrast, the present work addresses LQG control systems with noisy observations, where the state, observation, and action spaces are all continuous.

The work in \cite{postchannel} studies a class of finite-state channels in which the previous output serves as the current channel state, referred to as POST channels. The authors present a surprising result: for certain POST channels, feedback does not increase channel capacity. This work is relevant to ours because, to some extent, the implicit communication channel in an LQG control system with noiseless observations can also be viewed as a POST channel. At each time step, the system is in some state $x_t$, which can be interpreted as the channel state. The transmitter (i.e., the controller) applies an input $u_t$, resulting a new system state $x_{t+1}$, which can be viewed as the channel input and output, respectively. The current output $x_{t+1}$ then becomes the channel state at the next time step. The key differences between the work in \cite{postchannel} and ours are twofold: (i) they focus on channels with finite state and input alphabets; and (ii) their setting does not impose any constraints, whereas our formulation includes a control performance constraint, which is effectively a joint constraint on both inputs and outputs.

At the intersection of control and communication, extensive research has been conducted on control under communication constraints \cite{Mitter2004-constraint,Mitter2004-noisey,sabag2022reducing}, which is also known as networked control \cite{gupta2009networked}. These studies typically assume the presence of an explicit communication channel---often between the sensor and the controller---to enable information exchange within the control loop. The focus of analysis has been on understanding how the communication constraint affects the control performance. For example, the work in  \cite{kostina2019rate} analyzes the trade-off between control cost and the information rate over an explicit channel from the sensor to the controller. In contrast, our study introduces implicit communication as a new direction in integrating control and communication. In this framework, no explicit communication channel is assumed; rather, the control system itself serves as the medium for information exchange. From an information-theoretic perspective, implicit communication can be viewed as communication under control constraints, effectively reversing the conventional lens of networked control.

The problem of implicit communication can also be considered from a control-theoretic perspective. The work in \cite{Charalambous2017} examines a class of control problems in which the optimal control policy is stochastic. Specifically, their objective is to maximize the directed information from the control input sequence to the state sequence, subject to a general cost constraint. When their framework is specialized to linear Gaussian systems with a quadratic cost on the state and input, it becomes equivalent to our implicit communication setting under noiseless observations. In this special case, \cite{Charalambous2017} formulates an optimization problem to compute the maximum directed information rate—effectively the implicit channel capacity—but does not derive a closed-form expression for it. Follow-up studies on this framework include \cite{Charalambous-CDC, Charalambos2018-2, charalambous2024signalling}. Notably, \cite{charalambous2024signalling} considers LQG systems in which the encoder and decoder share the same noisy observations. This differs from our noisy setting, where the encoder and decoder have distinct observations, and no feedback is available from the decoder to the encoder. However, since observations on both sides are correlated with the system state, the implicit channel effectively operates as a channel with noisy feedback. 

%In \cite{wanglearning}, explicit communication is defined as those use direct channels independent of the system dynamics. These studies mostly rely on the so-called theory-of mind approach, which

Another line of research relevant to our work is the capacity of Gaussian MIMO channels, which has been extensively studied in the literature \cite{cover2002gaussian,verdu1994general,kim2009feedback,brandenburg1974capacity}.
Our analysis shows that, under a linear Gaussian input policy---which is proved to be capacity-achieving when the controller and the receiver have noiseless observations---the implicit channel can be equivalently translated into a memoryless Gaussian MIMO channel without channel state. The capacity of such channels is known to admit a closed-form expression, as derived in \cite{telatar1999capacity} via the water-filling algorithm. Our achievability proof in this setting builds on this result, with additional analysis related to the control constraint.

In the more general setting where the controller and the receiver have noisy observations, the equivalent channel under the capacity-achieving input policy becomes a Gaussian MIMO channel with memory, in which the noise process forms a hidden Markov chain.
The perfect feedback capacity of such channels is studied in \cite{sabag2023feedback} and is characterized as the solution to a finite-dimensional convex optimization problem. However, the feedback is noisy in our setting. To the best of our knowledge, the capacity of this type of Gaussian MIMO channel with memory but without feedback has not been presented in the literature; only some simulation-based algorithms for computing the information rate are available \cite{arnold2006simulation,feutrill2021review}.  It is also worth noting that the equivalent channel in our problem does not fall into the category of Gaussian channels with inter-symbol interference (ISI) \cite{hirt2002capacity,loyka2022capacity,Shamai1991}, as the memory arises solely from the hidden Markov structure of the noise process.

Overall, since it lies at the intersection of control and communication, implicit communication in LQG control systems can naturally be approached from both control-theoretic and information-theoretic perspectives. However, motivated by potential applications in practical systems, we are interested in two key questions: (i) the relationship between the capacity-achieving input policy for implicit communication and the optimal LQG control policy, and (ii) the trade-off between communication capacity and control cost. These considerations make implicit communication a compelling and fundamentally rich problem—and they form the core focus of this paper.

\section{System Model and Preliminaries}  \label{sec:system}
This section begins with an introduction to linear quadratic-Gaussian control systems. We then define implicit communication in this setting, illustrating how communication can occur as a byproduct of LQG control. Finally, we formulate the trade-off between communication and control performance.

\subsection{Linear Quadratic-Gaussian (LQG) Control Systems}
Consider a discrete-time linear control system
\begin{align}  \label{eq:LQG}
	x_{t+1}=Ax_t+Bu_t+w_t,
\end{align}
where $x_t\in \mathbb{R}^{d_1}$ and $u_t\in \mathbb{R}^{d_2}$ are the state and control input at time $t$, respectively, $w_t\in \mathbb{R}^{d_1}$ is an additive zero-mean white Gaussian noise with covariance matrix $\Psi_w\succ0$, and $A\in \mathbb{R}^{d_1\times d_1}$ and $B\in \mathbb{R}^{d_1\times d_2}$ are fixed matrices. We assume the noise is independent of the state and input. In addition, $\{w_t\}$ is an independent and identically distributed (i.i.d.) sequence, i.e., $w_t$ is independent of $w_k$ for any $t\neq k$. The initial state $x_1$ is randomly sampled from a Gaussian distribution $ \mathcal{N}(0,\Psi_x)$. Finally, we make the usual assumption that the pair $(A, B)$ is controllable.

At each time $t$, the controller observes the state $x_t$ and computes an input $u_t$ to control the system. 
The objective of control over a time horizon of length $n$ is to minimize the average quadratic cost, defined as
\begin{align} \label{eq:Jn}
	\min_{\{u_t\}} \ J_n \triangleq \frac{1}{n}\mathbb{E}\left[ \sum_{t=1}^{n} \left(x^\top_tFx_t + u^\top_tGu_t\right) + x^\top_{n+1}Fx_{n+1} \right],
\end{align}
where $F\in \mathbb{R}^{d_1\times d_1}$ and $G\in \mathbb{R}^{d_2\times d_2}$ are PSD matrices. 
The system is referred to as an LQG control system because it has linear dynamics, a quadratic cost function, and a Gaussian process noise. We next introduce some well-known results on LQG control, which will be useful in the following sections.

Let $J^*_n:=\min J_n$ denote the minimum cost of the $n$-step LQG control problem. The optimal control policy that achieves $J^*_n$ is linear and takes the form $u_t = -K^{(n)}_tx_t$, where the feedback gain $K^{(n)}_t$ is given by
\begin{align} \label{eq:optKn}
	K^{(n)}_t=(G+B^\top \Gamma^{(n)}_t B)^{-1} B^\top \Gamma^{(n)}_t A.
\end{align}
Here, $\Gamma^{(n)}_t$ is a PSD matrix determined by the following Riccati difference equation that runs backward in time:
\begin{align} \label{eq:Gamma-n}
	\Gamma^{(n)}_t=F+A^\top(\Gamma^{(n)}_{t+1}-\Gamma^{(n)}_{t+1} B(G+B^\top \Gamma^{(n)}_{t+1} B)^{-1}B^\top \Gamma^{(n)}_{t+1})A, \ 1\le t\le n,
\end{align}
where $\Gamma^{(n)}_{n+1} = F$.
Under the optimal control policy, the minimum cost $J^*_n$ is given by
\begin{align}  \label{eq:J*n}
	J^*_n =\frac{1}{n} \left[ \tr(\Psi_x \Gamma^{(n)}_1 )  + \sum_{t=1}^{n}\tr(\Psi_w \Gamma^{(n)}_{t+1})\right]. 
\end{align}
We refer to  \eqref{eq:Jn} as the $n$-step LQG control problem when $n$ is finite. For an infinite time horizon, the control objective is to minimize the long-term average quadratic cost, defined as:
\begin{align} \label{eq:q-cost}
	J\triangleq\lim_{n\to \infty} J_n=\lim_{n\to \infty} \frac{1}{n}\mathbb{E} \left[ \sum_{t=1}^{n} (x^\top _tFx_t + u^\top _tGu_t)  \right].
\end{align}
Denote by $J^*=\min J$ the minimum cost of the infinite time LQG control. The optimal control policy that achieves $J^*$ is  stationary linear and takes the form $u_t = -Kx_t$, where the feedback gain is
\begin{align} \label{eq:optK}
	K=(G+B^\top \Gamma B)^{-1} B^\top \Gamma A.
\end{align}
Matrix $\Gamma$ is the solution to the discrete-time algebraic Riccati equation (DARE):
\begin{align} \label{eq:Gamma}
	\Gamma=F+A^\top(\Gamma-\Gamma B(G+B^\top \Gamma B)^{-1}B^\top \Gamma)A.
\end{align}
Since the system is time-invariant---i.e., $A,B,F$ and $G$ are constant matrices over time, $(A,B)$ is controllable, and $\{w_t\}$ is a stationary noise process---there is a well-established relationship between finite-horizon and infinite-horizon LQG control problems. Specifically, as $n\to \infty$, the finite-horizon optimal cost converges to the infinite-horizon optimal cost, i.e., $J^*_n\to J^*$. Moreover, the feedback gain and Riccati solution stabilize, such that $\lim_{n\to \infty}K^{(n)}_t = K, \lim_{n\to\infty}\Gamma^{(n)}_t = \Gamma$ for all $t$. It follows immediately that the minimum average cost of the infinite-horizon control is given by  $J^* = \tr(\Gamma \Psi_w)$. 

\subsection{Implicit Communication in LQG Control Systems}
This paper studies implicit communication in LQG control systems, where the controller aims not only to minimize the average quadratic cost but also to transmit messages to a receiver through its control inputs. We assume there is no dedicated communication channel between the controller and the receiver; instead, both parties can observe the system state. As such, the LQG control system itself can serve as a communication channel between the controller and receiver, which we refer to as the \textit{implicit channel}. Communication via this implicit channel is called \textit{implicit communication} because (i) no explicit channel exists between the transmitter (i.e., the controller) and the receiver, and (ii) information transfer occurs as a ``byproduct'' of specifically designed control inputs.

\begin{figure}[t]
	\centering
	\includegraphics[width=0.8\linewidth]{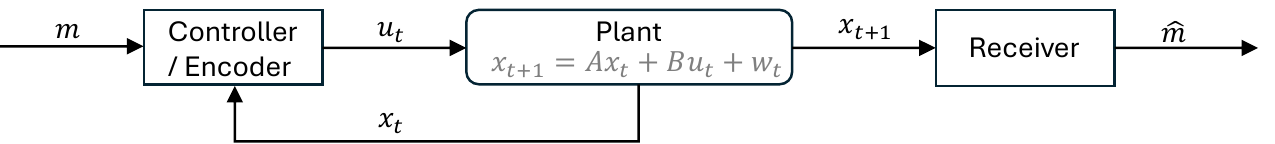}
	\caption{Implicit communication in LQG control systems (noiseless observations at controller and receiver).}
	\label{fig:model1}
\end{figure}

We begin with a simple setting in which both the controller and receiver have perfect access to the system state, as shown in Fig. \ref{fig:model1}. Extensions to more general scenarios, where observations are noisy for the receiver or for both parties, will be defined and addressed in later sections. 
To transmit a message $m$, the controller acts as an encoder, selecting an input $u_t$ at each time $t$ according to the message $m$ and the sequence of past and current states $x^t$. The receiver then observes the state sequences to decode the message $m$. Suppose the message set $\mathcal{M}=[2^{nR}]$ and each message $m$ is drawn uniformly from $\mathcal{M}$. 
Following standard definitions, a $(2^{nR},n)$ code for the implicit channel comprises a pair of  encoding and decoding mappings, used to transmit a message over $n$ time steps. In particular, let $\mathcal{U}=\mathbb{R}^{d_2}$ and $\mathcal{X}=\mathbb{R}^{d_1}$ denote the input and output alphabets, respectively. Then the encoder's mapping at time $t$ is denoted by
\begin{align*}
	u_t: \mathcal{M} \times \mathcal{X}^t \to \mathcal{U}.
\end{align*}
The decoder observes the initial state $x_1$ and the following $n$ states to decode the message, and uses the mapping 
\begin{align*}
	\hat{m}: \mathcal{X}^{n+1}\to \mathcal{M}.
\end{align*}
The average probability of error is then defined as $P^{(n)}_e = P(\hat{m}\neq m)$. Effectively, we treat the LQG control system as a channel with state: at each time $t$, an input $u_t$ is applied in state $x_t$, resulting an output $x_{t+1}$, which then becomes the channel state at the next time step. 

In general, communicating through the implicit channel requires the controller to deviate from the optimal control policy, resulting in a degradation of control performance. This leads to a fundamental trade-off between control and communication. A natural way to characterize this trade-off is by maximizing communication performance subject to a constraint on control cost. In particular, we say a $(2^{nR},n)$ code is admissible if it satisfies the control constraint
\begin{align*}
	J_n\le J^*_n + V, 
\end{align*}
where $J^*_n$ is the optimal $n$-step control cost defined in \eqref{eq:J*n} and $V\ge 0$ quantifies the tolerable control loss.
A rate $R$ is said to be achievable if there exists a sequence of admissible $(2^{nR},n)$ codes with $P^{(n)}_e\to 0$ as $n\to \infty$. The implicit channel capacity $C(V)$ is then defined as the supremum of all achievable rates under the control constraint.

\section{Noiseless Observations at the Controller and  the Receiver} \label{sec:perfect}
This section presents the capacity of the implicit channel under a control constraint. Given that perfect state observations are available at both the controller and the receiver, the capacity admits a closed-form expression that coincides with a memoryless Gaussian MIMO channel. In the special case of scalar systems, capacity $C(V)$ reduces to an explicit function of the minimum control cost $J^*$ and the tolerable control loss $V$. We provide the achievability proof in this section, as it offers useful insights into the structure of the channel under the capacity-achieving input policy. The converse proof is lengthy and is therefore deferred to Section \ref{sec: proofs}. 

\subsection{Capacity}
When both the controller and the receiver have noiseless state observations, the only source of noise in the implicit channel is the process noise $w_t$. As a result, the noise covariance matrix $\Psi_w$ plays a central role---not only in control, where the optimal long-term average control cost is given by $J^*=\tr(\Gamma \Psi_w)$, but also in communication. Since $\Psi_w$ is PD, $B^\top \Psi_w^{-1}B$ must be PSD. Therefore, $B^\top \Psi_w^{-1}B$ admits a diagonal decomposition of the from $B^\top \Psi_w^{-1}B=U\Lambda U^\top$, where $U\in \mathbb{R}^{d_2\times d_2}$ is a unitary matrix and $\Lambda=\mathtt{diag}(\lambda_1,\lambda_2,\cdots,\lambda_{d_2})$ is a diagonal matrix of eigenvalues. Suppose $B^\top \Psi_w^{-1}B$ has rank $r\le d_2$. Then $\Lambda$ contains $r$ positive diagonal entries and $d_2-r$ zero diagonal entries. Without loss of generality, we assume $U$ is chosen such that $\lambda_i>0$ for $i\in [r]$ and $\lambda_i=0$ for $r<i\le d_2$. We first highlight a key property of a matrix that is related to the capacity. 
\begin{lemma} \label{lem:hatGamma}
	For the LQG control system \eqref{eq:LQG} with noise covariance matrix $\Psi_w$, and suppose $B^\top\Psi^{-1}_w B=U\Lambda U^\top$ has rank $r$, as defined above. Define a matrix $\hat{\Gamma}\triangleq U^\top[B^\top \Gamma B+{G}] U$. Then $\hat{\Gamma}$ has non-negative diagonal entries, i.e., $\hat{\Gamma}(i,i)\ge 0$ for all $i$.
\end{lemma}

The proof of this lemma is straightforward. Since $\Gamma$ is a solution to the DARE defined in \eqref{eq:Gamma}, it is guaranteed to be PSD. By definition, $G$ is also PSD. Hence $\hat{\Gamma}$ must be PSD, which implies that all its diagonal entries are non-negative. Nevertheless, we provide an alternative and more insightful proof in the Appendix. This alternative argument shows if $\hat{\Gamma}(i,i)=0$ for some $i$, then it is possible to make the power of input $u_t$ unbounded in a certain direction without increasing the control cost. As will be elaborated later, this may lead to an unbounded capacity, i.e., $C(V) = \infty$. This special case can be ruled out by imposing a slightly stronger condition on $G$: if $G$ is PD, then $\hat{\Gamma}$ is also PD, ensuring that $\hat{\Gamma}(i,i)>0$ for all $i$.

We now present the capacity of the implicit channel when both the controller and the receiver have noiseless observations.
\begin{theorem} \label{thm:capacity-vec}
	For any $V\ge 0$, the capacity of the implicit channel under the control constraint $J\le J^* + V$ is given by
	\begin{align*}
		C(V) = \frac{1}{2} \sum_{i=1}^{r}\log  \left(1 + {\phi_i}{\lambda_i} \right),
	\end{align*}
	where ${\phi}_i=\infty$ if $\hat{\Gamma}(i,i)=0$ for $i\in [r]$; otherwise,
	\begin{align*}
		{\phi}_i= \max\left(\frac{\alpha}{\hat{\Gamma}(i,i)} -\frac{1}{\lambda_i},0\right),
	\end{align*}
	with $\alpha$ being chosen to satisfy
	\begin{align*}
		\sum_{i:i\le r,\hat{\Gamma}(i,i)>0} \max\left({\alpha}- \frac{\hat{\Gamma}(i,i)}{\lambda_i} ,0\right) = V.
	\end{align*}
	The optimal input that achieves the capacity is $u_t=-Kx_t + s_t$, where $K$ is the optimal feedback gain defined in \eqref{eq:optK} and $s_t\sim \mathcal{N}(0,\Phi)$ is independent of $w_t$ and $x_t$, with $\Phi=U\hat{\Phi} U^\top$ and $\hat{\Phi} = \mathtt{diag}(\phi_1,\phi_2,\cdots,\phi_{r},0,\cdots,0)$.
\end{theorem}

Theorem \ref{thm:capacity-vec} provides key insights into implicit communication in LQG control systems\footnote{{After posting this paper on arXiv, we were informed that a similar result was reported in \cite{Charalambous2017}, which investigates the maximization of the directed information rate from input to output under a general cost constraint. In the special case of LQG systems, their formulation reduces to the setting considered in Theorem 1 of our paper. The authors identified the linear Gaussian structure of the optimal policy and derived an equivalent optimization problem—using a different approach—but did not provide a closed-form expression for the implicit channel capacity.}}. First, the expression of the capacity matches that of Gaussian MIMO channels. As will be demonstrated clearly in the achievability proof, under the capacity-achieving input policy, the implicit channel effectively reduces to a memoryless Gaussian MIMO channel of the form $y_t = Bs_t + w_t$.

Second, Theorem \ref{thm:capacity-vec} reveals a \textit{separation principle}: the control input and the communication input can be designed independently, without loss of optimality. Specifically, the capacity-achieving input consists of two components: the term $-Kx_t$, which is exactly the optimal control input for minimizing the quadratic control cost, and the term $s_t$, a Gaussian random variable independent of $-Kx_t$. While the first term is fully determined by the control objective, communication is achieved by encoding information into the second term. Hence, we refer to $s_t$ as the \textit{communication signal}. As a consequence of this separation structure, given a tolerable control loss $V$, the task of maximizing communication performance under the control constraint can be decomposed into two subproblems: (i) computing the optimal LQG control policy; and then (ii) designing the channel code for a Gaussian MIMO channel. This decomposition greatly simplifies the implicit communication problem, as both subproblems are well studied and solvable using existing methods.

From a control perspective, the communication signal $s_t$ can be viewed as an additional noise injected into the process by the controller, thereby incurring an extra control cost. The exact value of this additional cost is given in the following lemma:

\begin{lemma}  \label{lem:constraint}
	For the LQG control system \eqref{eq:LQG}, under the input policy $u_t=-Kx_t + s_t$, where $s_t\sim \mathcal{N}(0,\Phi)$ is independent of $x_t$ and $w_t$, the long-term average control cost is given by $J=J^* + \tr((B^\top\Gamma B+{G}) \Phi)$.
\end{lemma}
\begin{IEEEproof} 
	See Appendix A.
\end{IEEEproof}

By the definition in \eqref{eq:q-cost}, the quadratic control cost comprises two components: a penalty for the state's deviation from 0 and an input cost. The communication signal not only raises the input cost by $\tr({G}\Phi)$, but also increases the state covariance in the steady-state, resulting in an additional cost of $\tr(\Gamma B \Phi B^\top)$.

Theorem \ref{thm:capacity-vec} can be specialized to scalar systems, resulting in a capacity expression that is an explicit function of the optimal control cost $J^*$ and tolerable control loss $V$.

\begin{theorem}  \label{thm:cap-scalar}
	For a scalar LQG system (i.e., $d_1=d_2=1$), the capacity of the implicit channel under the control constraint $J\le J^* + V$ is given by
	\begin{align*}
		C(V) =   \frac{1}{2}\log \left(1+ \frac{V}{J^*+B^{-2}G\Psi_w} \right).
	\end{align*}
	The optimal input is $u_t=-Kx_t + s_t$, where $K$ is the optimal feedback gain defined in \eqref{eq:optK} and $s_t\sim \mathcal{N}(0,V/(B^2\Gamma+G))$ is independent of $w_t$ and $x_t$.
\end{theorem}
\begin{IEEEproof}
	See Appendix A.
\end{IEEEproof}
 
 The expression in Theorem \ref{thm:cap-scalar} can be interpreted as follows. As discussed above, the term $B^{-2}G\Psi_w$ is equal to the input cost of a signal $B^{-1}w_t$—as if the process noise were injected by the control input itself. Therefore, the quantity $J^* + B^{-2}G\Psi_w$ can be viewed as the overall control cost when the noise $w_t$ originates from the controller rather than the environment. The capacity is then determined by the ratio of control costs in two scenarios: (i) the controller injects both $w_t$ and $s_t$ into the process, and (ii) the controller injects only $w_t$.

\subsection{Achievability Proof}
We now present the achievability proof for Theorem \ref{thm:capacity-vec}, showing that any rate $R\le C(V)$ can be achieved by the input policy $u_t=-Kx_t + s_t$, where $s_t\sim \mathcal{N}(0,\Phi)$ is independent of $x_t$ and $w_t$. Under this policy, the state evolves as follows:
\begin{align} \label{eq:th2-1}
	x_{t+1} = (A-BK)x_t + Bs_t + w_t.
\end{align}
Define $y_t \triangleq x_{t+1} - (A-BK)x_t$, then the above process becomes 
\begin{align} \label{eq:GaussianChan-1}
	y_t = Bs_t + w_t.
\end{align}
This forms a classic Gaussian MIMO channel with input $s_t$, output $y_t$, and additive white Gaussian noise $w_t$. The capacity of such a channel under an average input power constraint is well established. Hence the key step in our achievability proof is to show that, under the proposed policy, the control constraint can be equivalently translated into a constraint on the power of signal $s_t$. This result is established in Lemma \ref{lem:constraint}.
In particular, under the input policy $u_t=-Kx_t + s_t$,  the constraint $J\le J^*+V$ is equivalent to a weighted power constraint on $s_t$: $\tr((B^\top\Gamma B+{G}) \Phi) = \mathbb{E}[s^\top(B^\top\Gamma B+{G})s]\le V$. 
	
	The remainder of the proof largely follows the standard procedure for Gaussian MIMO channels \cite{telatar1999capacity}, with additional considerations related to the input power constraint, as matrix $B^\top\Gamma B+{G}$ is not guaranteed to be positive definite.
	Specifically, we use the well-known result that the capacity of the equivalent Gaussian MIMO channel defined in \eqref{eq:GaussianChan-1} is given by the maximum mutual information $\mathcal{I}(s;y)$, subject to the input constraint. Note that
	\begin{align*}
		 \mathcal{I}(s;y) =  h(y) - h(w) &=\frac{1}{2} \log \frac{\det(\Psi_w + B\Phi B^\top) }{\det(\Psi_w)} \\
         &=\frac{1}{2} \log \det(I + \Psi_w^{-\frac{1}{2}} B \Phi B^\top \Psi_w^{-\frac{1}{2}}) \\
         &= \frac{1}{2} \log \det(I + \Phi B^\top \Psi_w^{-1}B) \\
         & =\frac{1}{2} \log \det(I + \Phi U \Lambda U^\top ) \\
         &= \frac{1}{2} \log \det(I + \Lambda^{\frac{1}{2}} U^\top \Phi U \Lambda^{\frac{1}{2}}),
	\end{align*}
	where $\Lambda=\mathtt{diag}(\lambda_1,\lambda_2,\cdots,\lambda_{d_2})$. Recall that we assume $\Lambda$ has $r$ positive diagonal entries and $\lambda_i>0$ for $i\in [r]$.
	Let $\hat{\Phi} = U^\top \Phi U$ and $\hat{\Gamma} = U^\top (B^\top \Gamma B+{G}) U$, then the control constraint becomes
    $$\tr((B^\top \Gamma B+{G})  \Phi) = \tr((B^\top \Gamma B+{G})  U \hat{\Phi} U^\top) = \tr(\hat{\Gamma} \hat{\Phi})\le V.$$ Therefore, the capacity of the channel in \eqref{eq:GaussianChan-1} is given by
	\begin{align} \label{eq:CV-ach}
		C(V) = \max_{\hat{\Phi}\ge 0}& \ \frac{1}{2}\log \det(I + \Lambda^{\frac{1}{2}} \hat{\Phi} \Lambda^{\frac{1}{2}})  \\  \label{eq:CV-ach-st}
		\text{s.t.} & \ \tr(\hat{\Gamma} \hat{\Phi}) \le V.
	\end{align}
	It is known that the optimal $\hat{\Phi}$ is a diagonal matrix, because
	\begin{align*}
		\det(I + \Lambda^{\frac{1}{2}} \hat{\Phi} \Lambda^{\frac{1}{2}}) \le \prod_{i=1}^{d_2}\left(1+ \lambda_i\hat{\Phi}(i,i)\right)=\prod_{i=1}^{r}\left(1+ \lambda_i\hat{\Phi}(i,i)\right),
	\end{align*}
	where the inequality satisfies with equality when $\hat{\Phi}$ is diagonal.
    According to Lemma \ref{lem:hatGamma}, the diagonal entries of $\hat{\Gamma}$ are non-negative. We consider two scenarios: (1) $\hat{\Gamma}(i,i)>0$ for all $i\in [r]$; and (2) $\hat{\Gamma}(i,i)=0$ for some $i\in [r]$. 
    
    If it is case (1), the above problem can be solved by the well-known water-filling algorithm. In particular, the optimal $\hat{\Phi}$ is a diagonal matrix with the $i$-th diagonal entry given by
	\begin{align*}
		\hat{\Phi}(i,i)= \begin{cases}
		    \max\left(\frac{\alpha}{\hat{\Gamma}(i,i)} -\frac{1}{\lambda_i},0\right)  & \text{ if } i\in [r] \\
            0, &\text{ otherwise},
		\end{cases} 
	\end{align*}
	where ${\hat{\Gamma}(i,i)} $ is the $i$-th diagonal entry of $\hat{\Gamma}$, and $\alpha$ is chosen to satisfy
	\begin{align*}
		\sum_{i=1}^{r}\max\left({\alpha}- \frac{\hat{\Gamma}(i,i)}{\lambda_i} ,0\right) = V.
	\end{align*}
	Hence, the capacity is given by
	\begin{align*}
		{C}(V)  = \frac{1}{2} \sum_{i=1}^{r}\log  \left(1 + {\lambda_i } {\hat{\Phi}(i,i)}\right). 
	\end{align*}
	
	On the other hand, if $\hat{\Gamma}(i,i)=0$ for some $i\in [r]$, the channel capacity is infinite. Without loss of generality, assume that the first $k$ diagonal entries of $\hat{\Gamma}$ are zero ($k\le r$), and $\hat{\Gamma}(i,i)>0$ for $k<i\le r$. Then for diagonal $\hat{\Phi}$, the constraint can be written as
	\begin{align*}
		\tr(\hat{\Gamma} \hat{\Phi}) = \sum_{i=1}^{r} \hat{\Gamma}(i,i) \hat{\Phi}(i,i) = \sum_{i=k+1}^{r} \hat{\Gamma}(i,i) \hat{\Phi}(i,i)  \le V.
	\end{align*}
	In other words, if $\hat{\Gamma}(i,i)=0$, then there is no constraint on $\hat{\Phi}(i,i)$. We thus can make $C(V)$ unbounded by choosing  $\hat{\Phi}(i,i)=\infty$. Once this occurs, the values of $\hat{\Phi}(i,i)$ for $k<i\le r$ become irrelevant, as the capacity is already infinite. Nevertheless, for completeness and consistency, these remaining values can still be selected according to the water-filling principle. The only difference is that $\alpha$ is now chosen to satisfy
	\begin{align*}
		\sum_{i=k+1}^{r}\max\left({\alpha}- \frac{\hat{\Gamma}(i,i)}{\lambda_i} ,0\right) = V.
	\end{align*}
	This completes the achievability proof.

	By restricting attention to linear input policies, the achievability proof is significantly simplified via the channel translation. In contrast, the converse proof is more involved, as general input policies may render the system neither linear nor Gaussian. Fortunately, it can be shown that the mutual information $\mathcal{I}(m;x^{n+1})$ is maximized by linear Gaussian policies. On this basis, we prove the converse for the capacity theorem by Fano's inequality. Please refer to Section \ref{subsec:proof-thm1} for the converse proof.

%----------------------------------Section: --------------------------------------------------------------------
\section{Noisy Observations at Controller and Receiver} \label{sec:no}
\begin{figure}[t]
	\centering
	\includegraphics[width=0.9\linewidth]{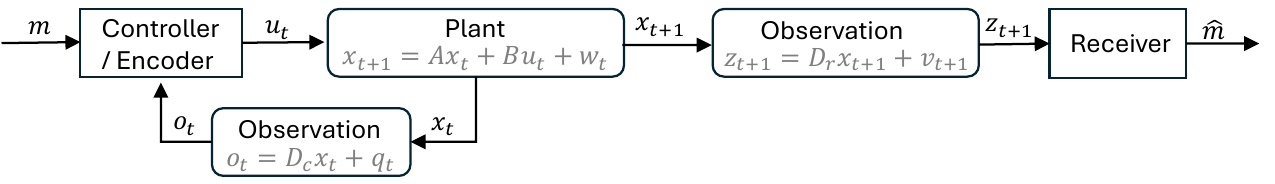}
	\caption{Implicit communication in LQG control systems with noisy observations at controller and receiver.}
	\label{fig:model3}
\end{figure}

In this section, we extend the implicit communication problem introduced in Section \ref{sec:system} to a general setting in which the controller and the receiver have noisy observations of the system state.  The system model is illustrated in  Fig. \ref{fig:model3}. Specifically, for the linear system defined in \eqref{eq:LQG}, the receiver has access to an observation given by
\begin{align} \label{eq:vt}
	z_t = D_rx_t + v_t,
\end{align}
where $v_t\in \mathbb{R}^{d_1}$ is zero-mean white Gaussian noise with covariance matrix $\Psi_v\succ 0$, and $D_r\in \mathbb{R}^{d_1\times d_1}$ is the observation matrix. We assume that $\{v_t\}$ is an i.i.d. process, independent of both the process noise $\{w_t\}$ and state sequence $\{x_t\}$. Additionally, we assume $D_r$ is invertible and the pair $(A,D_r)$ is observable. 
On the other hand, the controller observes $o_t\in \mathbb{R}^{d_3}$ given by
\begin{align} \label{eq:ot}
	o_t = D_c x_t + q_t, 
\end{align} 
where  $D_c\in \mathbb{R}^{d_3\times d_1}$ is the observation matrix of the controller, $q_t\sim \mathcal{N}(0,\Psi_q)$ is an observation noise independent of $\{x_i\}$, $\{v_i\}$ and $\{w_i\}$. We assume $\Psi_q\succ 0$, and that the pair $(A,D_c)$ is observable.

In this setting, the system can still function as an implicit channel between the controller and receiver. The difference is that the controller now computes input $u_t$ based on its observation history $o^t$, rather than the state sequence $x^{t}$. Therefore, the encoding mapping at time $t$ is denoted by
\begin{align*}
	u_t:\mathcal{M}\times \mathcal{O}^t \to \mathcal{U},
\end{align*}
where $\mathcal{O}=\mathbb{R}^{d_3}$ denotes the observation alphabet. Accordingly, the receiver decodes the message using the mapping:
\begin{align*}
	\hat{m}: \mathcal{Z}^{n+1} \to \mathcal{M},
\end{align*}
where $\mathcal{Z} = \mathbb{R}^{d_1}$.
The definitions of average probability of error and coding rate follow the standard conventions outlined in Section \ref{sec:system}. We denote by $C_{\text{no}}(V)$ the capacity of this implicit channel under a control constraint $J\le J^{**} + V$, where $J^{**}$ is the optimal control cost when the controller's observations are noisy, as will be discussed shortly.

Noisy observations at the controller and the receiver introduce substantial challenges in deriving the capacity of the implicit channel. Although we assume no feedback from the decoder to the encoder, the correlation between the controller’s and receiver’s observations means that the implicit channel effectively operates with noisy feedback and memory—a setting for which capacity characterization remains a long-standing open problem. 
As a result, we can only establish a lower bound on the capacity, which is achievable using linear Gaussian input policies.
Before presenting this bound, we first provide a brief discussion of control and state estimation under noisy controller observations, as some related quantities are needed in subsequent derivations.

\subsection{Optimal Control and Kalman Filtering}
Without considering implicit communication, the optimal control policy that minimizes the long-term average quadratic cost defined in \eqref{eq:q-cost} is given by $u_t = -K \check{x}_{t|t}$, where $K$ is the same feedback gain as in the perfect observation case, as defined in \eqref{eq:optK}, and $\check{x}_{t|t} \triangleq \mathbb{E}[x_t|o^t, u^{t-1}]$ is the controller’s estimate of the state at time $t$ based on its observations and past inputs. 
The estimate $\check{x}_{t|t}$ can be computed using the Kalman filter. In particular, define 
$$\check{x}_{t|k}\triangleq \mathbb{E}[x_t|o^k, u^{k-1}],$$ 
and let $\Sigma_{t|k} = \cov(x_t - \check{x}_{t|k})$ denote the corresponding estimation error covariance. The Kalman filter computes $\check{x}_{t|t}$ and $\Sigma_{t|t}$ recursively in two stages. In the prediction stage, the filter computes a prediction of the state  and its associated error covariance:
\begin{align*}
	\check{x}_{t+1|t} &= {A} \check{x}_{t|t} + Bu_t, \\
	\Sigma_{t+1|t} &= {A} \Sigma_{t|t}{A}^\top + \Psi_{{w}}.
\end{align*}
Upon receiving the new observation $o_{t+1}$, the state estimate is updated as:
\begin{align} \label{eq:nc-kf-2}
	\check{x}_{t+1|t+1} = \check{x}_{t+1|t} + L_{t+1} (o_{t+1} - D_c\check{x}_{t+1|t} ),
\end{align}
where $L_{t+1}$ is the filter gain given by
\begin{align*}
	L_{t+1} = \Sigma_{t+1|t} D_c^\top \left(D_c\Sigma_{t+1|t}D_c^\top  + \Psi_q \right)^{-1}.
\end{align*}
The estimation error covariance of $\check{x}_{t+1|t+1}$ is then given by
\begin{align*}
	\Sigma_{t+1|t+1}= (I-L_{t+1}D_c)\Sigma_{t+1|t}.
\end{align*}
Since the LQG control system is controllable and observable, it is a well-established result that both $\Sigma_{t|t}$ and $L_t$ converge. In particular, $\Sigma_{t+1|t}\to \Sigma_c$ as $t\to \infty$, where $\Sigma_c$ is the solution to the following discrete Riccati equation (see, e.g., \cite{krishnamurthy2016partially}):
\begin{align}
	\Sigma_c = A \left(\Sigma_c - \Sigma_c D_c^\top(D_c\Sigma_c D_c^\top + \Psi_q)^{-1}D_c\Sigma_c  \right)A^\top +  \Psi_{w}.
\end{align}
Note that $\Sigma_c$ is independent of the input policy, as the controller has full knowledge of the inputs.
Let $e_t \triangleq x_t - \check{x}_{t|t}$ denote the estimation error, we also have
\begin{align}
	& L_c\triangleq \lim_{t\to \infty}{L}_{t} = \Sigma_c  D_c^\top \left(D_c\Sigma_c D_c^\top  + \Psi_q \right)^{-1}, \\
	&\tilde{\Sigma}_c \triangleq \lim_{t\to\infty}\cov(e_t) = \lim_{t\to \infty} \Sigma_{t|t} =  (I-L_cD_c)\Sigma_c.
\end{align}
 Under the optimal policy $u_t = -K \hat{x}_{t|t}$, the minimum long-term average control cost is given by
\begin{align} \label{eq:opt-J-noise}
	J^{**} = \tr(F\tilde{\Sigma}_c) + \tr(\Gamma (\Sigma_c - \tilde{\Sigma}_c)),
\end{align}
where $\Gamma$ is defined in \eqref{eq:Gamma}.

\subsection{A Lower Bound on the Capacity}

In the setting with noiseless observations, the optimal input policy that achieves the implicit channel capacity adheres to a separation principle---the communication signal can be designed independently and then simply added to the optimal control input of the LQG control problem. This separation greatly simplifies the implicit communication problem, making it more practical and implementable. 
In the noisy observation setting, however, this separation principle no longer holds, as we will explain shortly.
Nevertheless, it is desirable for the input policy to retain a similar structure. Specifically, we focus on input policies of the form
\begin{align} \label{eq:LG-input}
	u_t  = -\bar{K}\check{x}_{t|t} + s_t, \quad s_t\sim \mathcal{N}(0,\Phi),
\end{align}
where $s_t$ is independent of $\check{x}_{t|t}$.
We refer to \eqref{eq:LG-input} as stationary linear Gaussian policies. 

In the context of implicit communication, linear Gaussian policies are arguably the most natural and practically meaningful choice. Given that the system is primarily designed for control, with communication as a secondary objective, it is important not to alter the fundamental structure of the control system. Linear Gaussian policies preserve the linear-Gaussian nature of the system dynamics, making them suitable for this purpose. Furthermore, as we will show in the next section, this class of policies enables an equivalent transformation of the implicit channel into a Gaussian MIMO channel, effectively decoupling the channel coding problem from the control design.

In this section, we derive the maximum achievable rate within this class of policies, which not only provides a lower bound on the implicit channel capacity but also carries significant practical importance, as linear Gaussian policies are a natural choice for real-world systems.
We first derive the control cost of the stationary linear Gaussian policy.
\begin{lemma} \label{lem:JforK}
    Given a policy $u_t = -\bar{K}\check{x}_{t|t} + s_t$, where $s_t\sim \mathcal{N}(0,\Phi)$ is independent of $\check{x}_{t|t}$, the long-term average control cost is given by 
    \begin{align*}
        J = J^{**} + \tr((B^\top \bar{\Gamma} B+ {G}) \Phi)  + \tr((\bar{\Gamma}-\Gamma) (\Sigma_c - \tilde{\Sigma}_c)), 
    \end{align*}
    where $\bar{\Gamma}$ is determined by the following equation:
    \begin{align} \label{eq:bargamma}
        \bar{\Gamma} = F+\bar{K}^\top G\bar{K} + (A-B\bar{K})^\top \bar{\Gamma} (A-B\bar{K}).
    \end{align}
\end{lemma}
\begin{IEEEproof}
    The proof is provided in Appendix B.
\end{IEEEproof}

Note that $\bar{\Gamma}$ is a constant matrix if $\bar{K}$ is fixed. In this case, the control cost constraint translates into a constraint on the power of the Gaussian signal $s_t$. Moreover, it is known that if $\bar{K}$ is chosen as the optimal LQG control gain defined in \eqref{eq:optK}, then equation \eqref{eq:bargamma} reduces to \eqref{eq:Gamma},  in which case $\bar{\Gamma}$ is equal to $\Gamma$.

Under the stationary linear Gaussian policy \eqref{eq:LG-input}, the system evolves according to the following equation:
\begin{align}  \label{eq:x-e-t}
	x_{t+1} = (A-B\bar{K})x_t + B\bar{K} e_t + Bs_t + {w}_t.
\end{align}
It is easy to verify that the error process $\{e_t\}$ is a Markov chain that evolves according to
\begin{align} \label{eq:et-controller}
	e_{t+1} = (I-L_cD_c){A}e_t + (I- L_cD_c)w_t - L_c q_{t+1} .
\end{align} 
While $e_t$ and $\check{x}_{t|t}$ are uncorrelated, we note that $e_t$ remains correlated with $x_t$:
\begin{align} \label{eq:correlation}
	 \cov(e_t,x_t)  = \mathbb{E}[e_t e^\top_t]  - \mathbb{E}[e_t \check{x}^\top_{t|t}] = \mathbb{E}[e_t e_t^\top] = \tilde{\Sigma}_c \succ 0. 
\end{align}
From the receiver's perspective, the controller's estimation error $e_t$ is an additional noise term introduced into the system. Because of the correlation between $e_t$ and $x_t$, the receiver cannot directly estimate $x_t$ using the standard Kalman filter, as it requires the noise to be White Gaussian and independent of the state $x_t$.

To address this issue, we construct an extended state $\rho_t\triangleq[x_t, e_t]$ and characterize the system from the receiver's perspective using the following equivalent form:
\begin{align} \label{eq:system-rho-1}
	\rho_{t+1} &= {A}_\rho \rho_t + \bar{s}_t + \bar{w}_t + \bar{q}_{t+1}, \\  \label{eq:system-rho-2}
	z_t & = D_\rho\rho_t + v_t,
\end{align}
where $D_\rho \triangleq [D_r\ 0]$ and
\begin{align} \label{eq:definitions}
	{A}_\rho \triangleq \begin{bmatrix}
		A-B\bar{K}& B\bar{K} \\
		0& (I-L_cD_c)A
	\end{bmatrix}, \ \bar{s}_t \triangleq \begin{bmatrix}
		Bs_t \\
		0
	\end{bmatrix}, \ \bar{w}_t \triangleq \begin{bmatrix}
		w_t \\
		(I-L_cD_c)w_t
	\end{bmatrix}, \ \bar{q}_{t+1} \triangleq \begin{bmatrix}
		0\\
		-L_c q_{t+1}
	\end{bmatrix}.
\end{align}
The system defined in \eqref{eq:system-rho-1}-\eqref{eq:system-rho-2} is equivalent to the one defined by \eqref{eq:x-e-t}-\eqref{eq:et-controller} and \eqref{eq:vt}.
Note that $\bar{s}_t, \bar{w}_t$, $\bar{q}_{t+1}$ and $v_t$ are all independent of each other, and all of them are independent of $\rho_t$---since $e_t$ is correlated only with $s^{t-1}, w^{t-1}$, and $q^t$. The covariance matrices of $\bar{s}_t, \bar{w}_t$, and $\bar{q}_{t+1}$ are given by
\begin{align} \label{eq: covariance}
	\bar{\Phi} = \begin{bmatrix}
		B\Phi B^\top & 0 \\
		0 & 0
	\end{bmatrix}, \Psi_{\bar{w}} = \begin{bmatrix}
		\Psi_w & \Psi_w (I-L_cD_c)^\top  \\
		(I-L_cD_c)\Psi_w & (I-L_cD_c)\Psi_w (I-L_cD_c)^\top
	\end{bmatrix}, \Psi_{\bar{q}} = \begin{bmatrix}
		0 & 0 \\
		0& L_c \Psi_q L_c^\top
	\end{bmatrix}.
\end{align}
Under this construction, the receiver can now estimate $\rho_t$ from its observations using the standard Kalman filter, setting the stage for deriving the lower bound.

\begin{theorem} \label{thm:lowerbound}
    For any $V\ge 0$, with noisy observations at the controller and the receiver, the capacity of the implicit channel under the control constraint $J\le J^{**} +V$ admits a lower bound $C_{\text{no}}(V) \ge C_{\text{no-lb}}(V)$, where $C_{\text{no-lb}}(V)$ is the maximum achievable rate over stationary linear Gaussian policies, obtained as the solution of the following optimization problem:
	\begin{align}
		C_{\text{no-lb}}(V) := \max_{\Phi, \bar{K}} \ & \frac{1}{2}\log {\det (D_\rho \Sigma_\rho D_\rho^\top + \Psi_v)} - \frac{1}{2}\log {\det(D_\rho \Pi D_\rho^\top + \Psi_v)} \\
		\text{s.t. } & \tr(({G}+B^\top \bar{\Gamma}B)\Phi )   + \tr((\bar{\Gamma} - \Gamma )(\Sigma_c - \tilde{\Sigma}_c)) \le V \\ \label{eq:hatgamma}
        & \bar{\Gamma} = F+\bar{K}^\top G\bar{K} + (A-B\bar{K})^\top \bar{\Gamma} (A-B\bar{K}) \\ \label{eq:sigma_rho}
		& \Sigma_\rho = {A}_\rho \left(\Sigma_\rho - \Sigma_\rho D_\rho^\top(D_\rho\Sigma_\rho D_\rho^\top + \Psi_v)^{-1}D_\rho\Sigma_\rho  \right){A}_\rho^\top + \bar{\Phi} + \Psi_{\bar{w}} +  \Psi_{\bar{q}} \\
        & \Phi \succeq 0,
	\end{align}
	where ${A}_\rho$ is defined in \eqref{eq:definitions}, $\bar{\Phi}, \Psi_{\bar{w}}$, and $ \Psi_{\bar{q}}$ are defined in \eqref{eq: covariance}, and $\Pi$ is given by the DARE:
	\begin{align} \label{eq:Pi}
    \Pi = {A}_\rho \left(\Pi - \Pi D_\rho^\top(D_\rho\Pi D_\rho^\top + \Psi_v)^{-1}D_\rho\Pi  \right){A}_\rho^\top + \Psi_{\bar{w}} +  \Psi_{\bar{q}}.
	\end{align}
\end{theorem}
\begin{IEEEproof}
    The proof is deferred to Section \ref{subsec:proof-thm3}
\end{IEEEproof}

Let $\hat{\rho}_{t+1|t}\triangleq\mathbb{E}[\rho_{t+1}|z^t]$ and $\bar{\rho}_{t+1|t}\triangleq\mathbb{E}[\rho_{t+1}|z^t,s^{t}]$ denote the optimal estimates of state $\rho_t$ based on $z^t$ and $(z^t,s^{t})$, respectively. Then $\Sigma_\rho$ and $\Pi$ are the corresponding steady-state estimation error covariances of $\hat{\rho}_{t+1|t}$ and $\bar{\rho}_{t+1|t}$. In the special case where both the controller and the receiver have noiseless observations (i.e., $D_r=D_c=I,\Psi_v=\Psi_q=0$), $\Sigma_\rho$ reduces to $B\Phi B^\top+\Psi_w$, and $\Pi$ reduces to $\Psi_w$. Moreover, $\bar{\Gamma}=\Gamma$ if $\bar{K}=K$. As a result,  the optimization problem in Theorem \ref{thm:lowerbound} simplifies to that underlying Theorem \ref{thm:capacity-vec}. In other words, this lower bound is tight in this special case. 

In general, the optimal $\bar{K}$ that achieves the lower bound $C_{\text{no-lb}}(V)$ is not equal to the optimal LQG control gain defined in \eqref{eq:optK}. This means that even when we restrict our attention to stationary linear Gaussian policies, the optimal feedback gain for implicit communication is, in general, different from that for standard LQG control—unlike in the noiseless setting, where the two coincide.
To see this more clearly, recall that in the noiseless setting, the implicit channel under a stationary linear Gaussian policy can be equivalently translated into a Gaussian MIMO channel:
\begin{align*}
    y_t = x_{t+1} - (A-B\bar{K})x_t = B s_t + w_t.
\end{align*}
Since the feedback gain is time-invariant and known to both the controller and the receiver, it does not affect the Gaussian channel directly. Therefore, the optimal feedback gain in this case is the one that minimizes the control cost, thereby allowing a larger power budget for the communication signal $s_t$.

In contrast, when observations are noisy, both the controller and the receiver must estimate the system state from their respective noisy observations. This introduces two sources of estimation error, both of which contribute to the overall channel noise. Since the feedback gain $\bar{K}$ influences both estimation errors, it plays a critical role in shaping the channel dynamics. Consequently, the optimal choice of $\bar{K}$ that achieves the maximum rate over stationary linear Gaussian policies must strike a careful balance between two competing objectives: (1) Minimizing the estimation error covariance to reduce the effective channel noise; (2) Minimizing the control cost, in order to maximize the power budget available for the communication signal. 
\begin{remark}
    In summary, when both the controller and the receiver have noiseless observations, the feedback gain influences only the input power constraint of the implicit channel. In contrast, in the noisy setting, the feedback gain affects both the input power constraint and the channel dynamics.
\end{remark}

Unfortunately, the optimization problem in Theorem \ref{thm:lowerbound} is not convex, and thus remains computationally onerous. However, the following result shows that for any fixed feedback gain $\bar{K}$, the optimization over $\Phi$ becomes a convex problem, which can be efficiently solved using standard methods \cite{cvx,gb08}.

\begin{proposition} \label{prop:f(K,V)}
    For any $V\ge 0$, the lower bound $C_{\text{no-lb}}(V)$ is given by 
    \begin{align*}
        C_{\text{no-lb}}(V) = \max_{\bar{K}} \ f(\bar{K},V), 
    \end{align*}
    where $f(\bar{K},V)$ is defined as the solution to the following convex optimization problem:
    \begin{align*}
        f(\bar{K},V): = \max_{\Phi,\Sigma_\rho} \ & \frac{1}{2}\log {\det (D_\rho \Sigma_\rho D_\rho^\top + \Psi_v)} - \frac{1}{2}\log {\det(D_\rho \Pi D_\rho^\top + \Psi_v)} \\
		\text{s.t. } & \tr(({G}+B^\top \bar{\Gamma} B)\Phi )  \le  V - \tr((\bar{\Gamma}-\Gamma) (\Sigma_c - \tilde{\Sigma}_c))  \\
		& \begin{bmatrix}
			D_\rho\Sigma_\rho D_\rho^\top + \Psi_v & D_\rho\Sigma_\rho {A}_\rho^\top \\
			{A}_\rho\Sigma_\rho D_\rho^\top & {A}_\rho\Sigma_\rho {A}_\rho^\top + \bar{\Phi} + \Psi_{\bar{w}} +  \Psi_{\bar{q}} - \Sigma_\rho
		\end{bmatrix} \succeq 0 \\
		& \Sigma_\rho \succeq 0, \Phi \succeq 0.
    \end{align*}
    where $\bar{\Gamma}$ is determined by $\bar{K}$ via the equation \eqref{eq:hatgamma}.
\end{proposition}
\begin{IEEEproof}
    See Appendix B.
\end{IEEEproof}

\subsection{Channel Translation}

As previously discussed, when both the controller and the receiver have noisy observations, the implicit channel becomes a memory channel with state and operates under noisy feedback. Designing practical coding schemes for such channels is notoriously challenging, posing a significant obstacle to the practical implementation of implicit communication. However, if we restrict our attention to stationary linear Gaussian policies, the implicit channel can be equivalently transformed into a Gaussian MIMO channel with memory—but without state information or feedback—making it much more amenable to existing coding techniques. This section presents this channel translation and shows that, for any fixed feedback gain $\bar{K}$, the capacity of the equivalent Gaussian channel is given by $f(\bar{K},V)$ defined in Proposition \ref{prop:f(K,V)}.

To construct a Gaussian channel that is equivalent to the implicit channel under the stationary linear Gaussian policy $u_t = -\bar{K}\check{x}_{t|t}+ s_t$, we focus on the extended system \eqref{eq:system-rho-1}-\eqref{eq:system-rho-2} and define
\begin{align} \label{eq:output-trans}
    y_t \triangleq \rho_{t+1} - A_\rho \rho_t=\bar{s}_t + \bar{w}_t + \bar{q}_{t+1}.  
\end{align}
The receiver cannot observe $y_t$ directly; instead, it must estimate $y_t$ from its observations.  In particular, let $\hat{\rho}_{t|k}\triangleq \mathbb{E}[\rho_t|z^k]$ denote the receiver's estimate of $\rho_t$ based on $z^k$, and $\Sigma_{t|k}$ the corresponding estimation error covariance. Then the receiver can estimate $y_t$ from $z^{t+1}$ as follows:
\begin{align} \label{eq:haty}
    \hat{y}_t \triangleq \mathbb{E}[y_t|z^{t+1}] = \hat{\rho}_{t+1|t+1} - A_\rho \hat{\rho}_{t|t+1}.
\end{align}
The first term, $\hat{\rho}_{t+1|t+1} $, can be computed using the standard Kalman filter. Unlike the controller, however, the receiver does not know the input, and therefore, treats $\bar{s}_t$ as an additional noise term.
The second term, $\hat{\rho}_{t|t+1}$, can be computed by the Kalman smoother (also known as the Rauch-Tung-Striebel smoother \cite{smoother}),  which provides smoothed estimates of past states using future observations. Specifically, the Kalman smoother computes
\begin{align} \label{eq:smoother-rho}
		\hat{\rho}_{t|t+1} = \hat{\rho}_{t|t} + Q_t \left(\hat{\rho}_{t+1|t+1} - {A}_\rho \hat{\rho}_{t|t}\right),
	\end{align}
where the smoothing gain $Q_t$ is defined as $Q_t \triangleq \Sigma_{t|t}{A}_\rho^\top \Sigma_{t+1|t}^{-1}$. For any $\bar{K}$ such that $A-B\bar{K}$ is stable, both the Kalman filter and smoother converge within a finite time. As a result, $\Sigma_{t+1|t}\to\Sigma_\rho$ as $t\to\infty$, where $\Sigma_\rho$ is determined by the DARE \eqref{eq:sigma_rho}. Accordingly, we also have
	\begin{align}  \label{eq: Q-inf-rho}
		&\lim_{t\to\infty} Q_t = Q_\rho \triangleq  (I-L_\rho D_\rho)\Sigma_\rho {A}_\rho^\top \Sigma_\rho^{-1}, \\
        & \lim_{t\to\infty}\Sigma_{t|t} = \tilde{\Sigma}_\rho \triangleq (I-L_\rho D_\rho)\Sigma_\rho,
	\end{align}
where
\begin{align*}
    L_\rho \triangleq \Sigma_\rho D_\rho^\top(D_\rho \Sigma_\rho D_\rho^\top + \Psi_v)^{-1}.
\end{align*}
Throughout this section, we assume that the Kalman filter and smoother have reached steady state. As a result, the estimation error $\tau_t \triangleq \rho_t - \hat{\rho}_{t|t}$ evolves as a stationary Markov process, given by
\begin{align*}
	\tau_{t+1} =  (I-L_\rho D_\rho)A_\rho \tau_t - L_\rho v_{t+1} + (I-L_\rho D_\rho)( \bar{s}_t + \bar{w}_t + \bar{q}_{t+1}). 
\end{align*}

Since estimating $y_t$ reduces to estimating $\rho_t$ and $\rho_{t+1}$, and the receiver has access to the entire observation sequence $z^{n+1}$ at the time of decoding, it may seem that the optimal estimate of $y_t$ is $\mathbb{E}[y_t|z^{n+1}]$ rather than $\mathbb{E}[y_t|z^{t+1}]$. If this were the case, then the resulting estimate of $y_t$ would depend not only on past inputs but also on future ones.
Fortunately, it turns out that the sequence $\hat{y}^n$ generated by the transformation \eqref{eq:haty} retains all information about $s^n$, with no loss. This key property allows us to translate the implicit channel into an equivalent Gaussian channel, as formalized below.

\begin{theorem} \label{thm: channel-trans}
	Consider the input policy $u_t = -\bar{K}\check{x}_{t|t} +  s_t$, where $s_t\sim \mathcal{N}(0,\Phi)$ is independent of $\check{x}_{t|t}$. The implicit channel under the control constraint $J\le J^{**} +V$ is equivalent to the Gaussian MIMO channel
	\begin{align*}
		\bar{y}_t = D_r Bs_t + \beta_t, 
	\end{align*}
	subject to the input power constraint $\mathbb{E}[s_t^\top(B^\top \bar{\Gamma} B + {G}) s_t]\le V- \tr((\bar{\Gamma}-\Gamma) (\Sigma_c - \tilde{\Sigma}_c))$, 
	where $\bar{\Gamma}$ is determined by $\bar{K}$ via \eqref{eq:hatgamma}, the channel output $\bar{y}_t$ is a function of $z^{t+1}$ given by
    \begin{align*}
        \bar{y}_t = L_\rho^\dagger (I-A_\rho Q_\rho)^{-1}\hat{y}_t,
    \end{align*}
    and the noise term is given by $\beta_t = D_rw_t +  D_\rho A_\rho \tau_t + v_{t+1}$. Moreover, the capacity of this channel is $f(\bar{K},V)$.
\end{theorem} 
\begin{IEEEproof}
	The proof is deferred to Section \ref{subsec:proof-thm4}.
\end{IEEEproof}

The noise process $\{\beta_t\}$ in the equivalent Gaussian channel is a hidden Markov process. Therefore, the estimation error $\tau_t$ can be effectively viewed as the hidden channel state, evolving according to a stationary Markov process. Designing practical coding schemes for such channels remains a challenging task. However, the fact that the equivalent channel operates without feedback and without channel state information at either the transmitter or receiver may help simplify the coding problem. More importantly, this channel translation decouples the channel coding task from control design once the feedback gain $\bar{K}$ is fixed, allowing the two components to be addressed independently, and thereby, significantly simplifying the overall problem.

It is insightful to further examine the noise term in the equivalent Gaussian channel. Specifically, let $\sigma_t = x_t - \mathbb{E}[x_t|z^t]$ and $\varepsilon_t = e_t - \mathbb{E}[e_t|z^t]$. Then $\tau_t = [\sigma_t, \varepsilon_t]$ and we can decompose $D_\rho A_\rho \tau_t$ as follows:
\begin{align*}
	D_\rho A_\rho \tau_t = D_r(A-B\bar{K}) \sigma_t + D_rB\bar{K} \varepsilon_t.
\end{align*}
Note that $\sigma_t$ arises from the receiver’s observation noise, while $\varepsilon_t$ originates from the controller’s observation noise. The expression above illustrates how the noises from both sides contribute to the overall channel noise. The correlation between $\sigma_t$ and $\varepsilon_t$ further complicates the overall noise structure. Nevertheless, by leveraging the extended linear system defined in \eqref{eq:system-rho-1}–\eqref{eq:system-rho-2}, these two correlated noise components can be jointly treated as a single process.

\section{Proofs} \label{sec: proofs}
This section begins with the converse proof for Theorem \ref{thm:capacity-vec}, and then presents the proofs of Theorems \ref{thm:lowerbound} and \ref{thm: channel-trans}. The converse proof is based on Fano's inequality and demonstrates that the mutual information between the message and the receiver's observation sequence is upper-bounded by the channel capacity. The central step is to demonstrate that this mutual information is maximized by input policies of the form $u_t = -Kx_{t} + s_t$.
However, in the setting where both the controller and the receiver have noisy observations, we are unable to establish a similar property—constituting the main obstacle to deriving the exact capacity in this case.

\subsection{Converse Proof of Theorem \ref{thm:capacity-vec}} \label{subsec:proof-thm1}
To prove the converse to Theorem \ref{thm:capacity-vec}, we show that a sequence of $(2^{nR_n},n)$ codes with $P_e^{(n)}\to 0$ must have $R_n \le \bar{C}_n(V) + \epsilon_n$, where $\epsilon_n\to 0$ and $\bar{C}_n(V)\to C(V)$.
\begin{IEEEproof}[Converse Proof]
	First of all, by Fano's inequality,
	\begin{align*}
		nR_n = H(m) & = H(m|x^{n+1}) + \mathcal{I}(m;x^{n+1})  \\
		& = \mathcal{I}(m;x^{n+1}) + n\epsilon_n,
	\end{align*}
	where $\epsilon_n\to 0$ if $P_e^{(n)}\to 0$. Now,
	\begin{align*}
		\mathcal{I}(m;x^{n+1}) & = h(x^{n+1}) - h(x^{n+1}|m) \\
		& = \sum_{i=2}^{n+1} h(x_i|x^{i-1}) + h(x_1) - \sum_{i=2}^{n+1} h(x_i|x^{i-1},m) - h(x_1|m) \\
		& = \sum_{i=1}^{n} h(x_{i+1}|x^{i}) - \sum_{i=1}^{n} h(x_{i+1}|x^{i},m).
	\end{align*}
	Here, $h(x_1) - h(x_1|m)=0$ because $x_1$ is the initial state, which is independent of the message $m$.
	Since $u_i$ is determined by $x^i$ and $m$, we have
	\begin{align*}
		h(x_{i+1}|x^{i},m) & = h(Ax_{i}+Bu_{i}+w_{i}|m,x^{i},u_{i}(x^{i},m)) \\
		& = h(w_{i}).		
	\end{align*}
	Also, $h(x_{i+1}|x^{i}) = h(Ax_{i}+Bu_{i}+w_{i}|x^{i}) = h(Bu_{i}+w_{i}|x^{i})$.
	We thus have
	\begin{align*}
		\mathcal{I}(m;x^{n+1}) & = \sum_{i=1}^{n} h(Bu_{i}+w_{i}|x^{i}) - \sum_{i=1}^{n} h(w_i)  \\
		& \le \sum_{i=1}^{n} h(Bu_{i}+w_{i}|x_i) - \frac{n}{2} \log (2\pi e)^{d_1} \det(\Psi_w).
	\end{align*}
	For $n\ge 1$, defined a sequence of optimization problems as follows:
	\begin{align}  \label{eq:cv-obj-ve}
		\bar{C}_n(V): = \max \ & \frac{1}{n} \sum_{i=1}^{n} \left[h(Bu_{i}+w_{i}|x_i) - \frac{1}{2} \log (2\pi e)^{d_1} \det(\Psi_w) \right] \\  \label{eq:cv-st-ve}
		\text{s.t.} \ & J_n\le J_n^* + V.
	\end{align}
	Then a sequence of $(2^{nR_n},n)$ codes satisfying constraint $J_n\le J^*_n +V$ and $P_e^{(n)}\to 0$ must have
	\begin{align*}
		R_n \le  \frac{1}{n}\mathcal{I}(m;x^{n+1}) + \epsilon_n \le \bar{C}_n(V)  + \epsilon_n,
	\end{align*} 
	where $\epsilon_n\to 0$ if $P_e^{(n)}\to 0$. It remains to show that $\bar{C}_n(V)\to C(V)$ as $n\to \infty$. The following lemma plays a central role in solving $\bar{C}_n(V)$:
	
	\begin{lemma}  \label{lem:opt-policy}
		For any $n\ge 1$, it is sufficient to optimize $\bar{C}_n(V)$, as defined in \eqref{eq:cv-obj-ve}-\eqref{eq:cv-st-ve}, over input policies of the form $u_t = -K^{(n)}_t x_t + s_t$, where $s_t\sim \mathcal{N}(0,\Phi_t)$ is independent of $x_t$ and $w_t$.
	\end{lemma}
	\begin{IEEEproof}
		 Proof of Lemma \ref{lem:opt-policy} is presented immediately after the converse proof.
	\end{IEEEproof}
	
	Lemma \ref{lem:opt-policy} implies that we can focus on $u_t = -K^{(n)}_t x_t + s_t$, $s_t\sim \mathcal{N}(0,\Phi_t)$, for solving the optimization problem. Under this policy,
	\begin{align*}
		h(Bu_t + w_t|x_t) = h(Bs_t + w_t) = \frac{1}{2} \log (2\pi e)^{d_1} \det(B\Phi_tB^\top + \Psi_w).
	\end{align*}
	We thus can convert problem \eqref{eq:cv-obj-ve}-\eqref{eq:cv-st-ve}  to an equivalent optimization problem as follows:
	\begin{align}  \label{eq:cv2-obj-vec}
		\bar{C}_n(V): = \max_{\{\Phi_t\ge 0\}} \ & \frac{1}{2n} \sum_{t=1}^{n} \left[  \log \det \left(\Psi_w+ B{\Phi_t}B^\top\right) - \log \det(\Psi_w) \right]  \\  \label{eq:cv2-st-vec}
		\text{s.t.} \ & \frac{1}{n} \sum_{t=1}^{n} \tr(\Phi_t ( B^\top \Gamma^{(n)}_{t+1} B+ {G})) \le  V.
	\end{align}
	Since $\Gamma^{(n)}_t = \Gamma$ for all $t$ as $n\to \infty$, it is easy to verify that, as $n\to \infty$, $\bar{C}_n(V)$ converges to the following problem:
	\begin{align}  \label{eq:cv2-obj-inf}
		\bar{C}_\infty(V) = \max_{\Phi\ge 0} \ & \frac{1}{2} \left[  \log \det \left(\Psi_w+ B{\Phi_t}B^\top\right) - \log \det(\Psi_w) \right]  \\  \label{eq:cv2-st-inf}
		\text{s.t.} \ & \tr(\Phi (B^\top \Gamma B+{G})) \le  V.
	\end{align}
	Using the same argument as in the achievability proof, the above problem reduces to the problem $C(V)$ defined in \eqref{eq:CV-ach}-\eqref{eq:CV-ach-st}. We thus conclude that
	\begin{align*}
		\lim_{n\to \infty}\bar{C}_n(V) =\bar{C}_\infty(V) =  C(V).
	\end{align*}
	This completes the converse proof.
\end{IEEEproof}

Next, we prove Lemma \ref{lem:opt-policy}, a key component of the converse proof.
	
\begin{IEEEproof} [Proof of Lemma \ref{lem:opt-policy}]
	In general,  $u_i$ is a deterministic function of historical states $x^i$ and message $m$. However, since $h(Bu_i+w_i|x_i)$ depends only on the conditional distribution of $u_i$ given $x_i$, we thus focus on $x_i$ and treat $u_i(x_i)$ as a stochastic function of $x_i$. By definition, the $n$-step average control cost can be expressed as
	\begin{align} \label{eq:Jn-vec}
		J_n & = \frac{1}{n}\sum_{t=1}^{n+1}\mathbb{E} [x^\top_tFx_t]  +  \frac{1}{n}\sum_{t=1}^{n} \mathbb{E}  [u^\top_tGu_t ] \notag  \\
		& = \frac{1}{n}\sum_{t=1}^{n+1} \left[\tr(F\Psi_{x_t}) + \mathbb{E}[x^\top_t]F\mathbb{E}[x_t ] \right] +  \frac{1}{n}\sum_{t=1}^{n} \left[ \tr(G\Psi_{u_t}) +  \mathbb{E}[u^\top_t]G\mathbb{E}[u_t ] \right],  
	\end{align}
	where $\Psi_{x_t}$ and $\Psi_{u_t}$ denote the covariance matrices of $x_t$ and $u_t$, respectively. Equation \eqref{eq:Jn-vec} suggests that a constraint on $J_n$ effectively translates into restrictions on the mean and covariance of states and inputs. It is known that, for a given covariance, the differential entropy is maximized by a Gaussian distribution. Therefore, problem \eqref{eq:cv-obj-ve}-\eqref{eq:cv-st-ve} can be optimized by a policy in which the conditional distribution of $u_t$ given $x_t$ is Gaussian. We refer to such policies as the type-1 policies and formally define them as follows: 
	
	\noindent\textbf{Type-1 Policies:} A policy $\tau$ is classified as a type-1 policy if, for any time $t$, the conditional distribution of $u_t$ given $x_t$ is a Gaussian distribution $ \mathcal{N}(\theta_t(x_t), \Psi_t(x_t))$, where both the mean $\theta_t$ and covariance matrix $\Psi_t$ are functions of the state $x_t$.
	
	Next, we introduce a subset of type-1 policies, referred to as type-2 policies, which are defined as follows:
	
	\noindent\textbf{Type-2 Policies:} A policy $\tau'$ is classified as a type-2 policy if, for any time $t$, the conditional distribution of $u_t$ given $x_t$ is a Gaussian distribution $ \mathcal{N}(\theta'_t(x_t), \Phi'_t)$, where the mean $\theta'_t$ is a function of the state $x_t$, while the covariance matrix $\Phi'_t$ is independent of $x_t$.
	
	Now that $\bar{C}_n(V)$ can be optimized by type-1 policies, we proceed to prove Lemma \ref{lem:opt-policy} in two steps: (i) in step 1, we show that it is sufficient to optimize $\bar{C}_n(V)$ over the type-2 policies; (ii) in step 2, we further demonstrate that any type-2 policy has an equivalent linear policy.
	
	In step 1, we begin by considering an arbitrary type-1 policy $\tau$ and an associated type-2 policy $\tau'$ satisfying $\theta'_t(x_t)=\theta_t(x_t)$ and $\Phi'_t =\Phi_t\triangleq \mathbb{E}[\Psi_{t}(x_t)]$ for all $t$. The goal is to show that both policies yield the same control cost, and that $\tau'$ achieves an objective value no lower than that of $\tau$.
	First, note that, for a fixed distribution of $x_t$, both policies induce the same marginal distribution of $u_t$. To see this, define $\rho_t \triangleq  \mathbb{E}[\theta_{t}(x_t)]$. Then clearly, under both policies, $\mathbb{E}[u_t] = \rho_t$. An application of the law of total covariance shows that the covariance of $u_t$ is also the same under both policies. In particular, the following holds for both policies: 
	\begin{align} \label{eq:LTV}
		\cov(u_t) = \mathbb{E}\left[\cov( u_t|x_t)\right] + \cov\left(\mathbb{E}[u_t|x_t]\right) = \Phi_t + \cov(\theta_{t}(x_t)).
	\end{align}
	Therefore, the marginal distribution of $u_t$ under both $\tau$ and $\tau'$ is $ \mathcal{N}(\rho_t, \Phi_t + \cov(\theta_{t}(x_t)))$. 
	
	We then demonstrate that  $\tau$ and $\tau'$ yield the same control cost for the $n$-step LQG control problem. According to \eqref{eq:Jn-vec}, this can be established by showing that, for any $t$, the distributions of $x_t$ under the two policies are identical. This claim can be proven using mathematical induction.  
	Initially, note that $x_1 \sim \mathcal{N}(0,\Psi_x)$ is independent of the control policy, hence the distributions of $x_1$ under both policies are identical. Assume as the induction hypothesis that the distributions of $x_t$ under the two policies are identical. 
	Now, 
	\begin{align*}
		x_{t+1} = A x_t + Bu_t(x_t) + w_t.
	\end{align*}
	Denote by $\mathcal{N}(\nu, \Psi)$ and $\mathcal{N}(\nu', \Psi')$ the marginal distributions of $A x_t + Bu_t$ under policies $\tau$ and $\tau'$, respectively. We have
	\begin{align*}
		\nu = A\mathbb{E}[x_t] + \mathbb{E}[\mathbb{E}[Bu_t(x_t)|x_t]] = A\mathbb{E}[x_t] + B\rho_t = \nu',
	\end{align*} 
	and
	\begin{align*}
		\Psi &= \mathbb{E}[(Ax_t +Bu_t)(Ax_t +Bu_t)^\top ]- \mathbb{E}[Ax_t + Bu_t] \mathbb{E}[Ax_t + Bu_t]^\top \\
		& = A \mathbb{E}[x_tx^\top_t]A^\top + \mathbb{E}[(Bu_t)(Bu_t)^\top] + A\mathbb{E}[x_tu^\top_tB^\top] +\mathbb{E}[Bu_tx^\top_t]A^\top   - \nu \nu^\top \\
		& = A \mathbb{E}[x_tx^\top_t]A^\top + \mathbb{E}[(Bu_t)(Bu_t)^\top] + A\mathbb{E}[\mathbb{E}[x_tu^\top_tB^\top|x_t]] +\mathbb{E}[\mathbb{E}[Bu_tx^\top_t|x_t]]A^\top   - \nu \nu^\top \\
		& = A \mathbb{E}[x_tx^\top_t]A^\top + \mathbb{E}[(Bu_t)(Bu_t)^\top] + A\mathbb{E}[x_t\theta_t(x_t)^\top B^\top] +\mathbb{E}[B\theta_t(x_t)x^\top_t]A^\top   - \nu \nu^\top \\
		& = \Psi'.
	\end{align*}
	The above implies that the distributions of $A x_t + Bu_t$ under both policies are identical.
	Since $x_t$ is independent of $w_t$, so is $u_t$. As a result, the marginal distributions of $x_{t+1}$ under policies $\tau$ and $\tau'$ are identical, completing the inductive step. We thus conclude that, starting from the same initial state distribution, policies $\tau$ and $\tau'$ produce identical state and (marginal) input distributions at every time step. According to \eqref{eq:Jn-vec}, two policies result in the same control cost if they produce identical state and  input distributions at every time step. Hence the control costs of $\tau$ and $\tau'$ are equal.
	
	Next, we show that the objective value of $\tau'$ is greater than or equal to that of $\tau$. As proved earlier, policies $\tau$ and $\tau'$ produce identical distributions of $x_t$,  $\forall t\ge 1$. Let $f_t(x)$ denote the probability density function of $x_t$. Under policy $\tau$,
	\begin{align*}
		h_t^\tau \triangleq h(Bu_t + w_t|x_t) =\frac{1}{2} \int_{-\infty}^{+\infty}f_t(x_t) \log (2\pi e)^{d_1} \det(B\Psi_t(x_t)B^\top + \Psi_w) dx_t.
	\end{align*} 
	Under policy $\tau'$, we can express $u_t(x_t) = \theta_t(x_t) + s_t$, where $s_t\sim \mathcal{N}(0, \Phi_t)$ is independent of $x_t$ and $w_t$. Then
	\begin{align*}
		h_t^{\tau'} \triangleq h(Bu_t + w_t|x_t) = h(Bs_t+w_t) = \frac{1}{2} \log (2\pi e)^{d_1} \det(B\Phi_tB^\top + \Psi_w).
	\end{align*}
	Since $\log \det(\cdot)$ is a concave function, Jensen's inequality implies: 
	\begin{align*}
		h_t^\tau &=\frac{1}{2} \mathbb{E} \left[ \log \det(B\Psi_t(x_t)B^\top + \Psi_w) \right] + \frac{d_1}{2}\log 2 \pi e  \\
		& \le \frac{1}{2}  \log \det(B\mathbb{E}[\Psi_t(x_t)]B^\top + \Psi_w) + \frac{d_1}{2}\log 2 \pi e =h_t^{\tau'}.
	\end{align*}
	It follows immediately that the objective value achieved by $\tau'$ is at least as large as that of $\tau$. Combining the above arguments together, the optimal value of the optimization problem \eqref{eq:cv-obj-ve}-\eqref{eq:cv-st-ve} can be attained by policies of the form $u_t = \theta_t(x_t) +s_t$, where $\theta_t(\cdot)$ is a deterministic function and $s_t\sim \mathcal{N}(0,\Phi_t)$ is independent of $x_t$ and $w_t$. 
	
	We then move on to step 2. We will show that, for an arbitrary policy of the form $u_t = \theta_t(x_t) + s_t$ that achieves an objective value $C_n$ and constraint value (i.e., control cost) $J_n$ in problem \eqref{eq:cv-obj-ve}-\eqref{eq:cv-st-ve}, there exists a linear Gaussian policy of the form $u_t = -K_t x_t + s_t$ that achieves the same objective value while attaining a constraint value of as most $J_n$. 
	
	In particular, let $u_t = \theta_t(x_t) + s_t$, where $s_t\sim \mathcal{N}(0,\Phi_t)$ is independent of both $x_t$ and $w_t$. By definition, the control cost of this policy for the $n$-step control problem is given by
	\begin{align*}
		J_n &= \frac{1}{n}\sum_{t=1}^{n+1}\mathbb{E} [x^\top_tFx_t ]  +  \frac{1}{n}\sum_{t=1}^{n} \mathbb{E}  [u^\top_tGu_t ] \notag  \\
		& =  \frac{1}{n}\sum_{t=1}^{n+1}\mathbb{E} \left[x^\top_tFx_t  \right]  + \frac{1}{n}\sum_{t=1}^{n}\mathbb{E} \left[ \theta_t(x_t)^\top G \theta_t(x_t)  \right]  +  \frac{1}{n}\sum_{t=1}^{n} \mathbb{E}  [ s_t^\top G s_t] \notag  \\
		& \triangleq J'_n + \frac{1}{n}\sum_{t=1}^{n} \tr(G\Phi_t).
	\end{align*}
	Under this policy, the system state evolves as $x_{t+1} = A x_t + B\theta_t(x_t) + Bs_t + w_t$. By treating $Bs_t$ as an additional noise introduced into the system, we can interpret $J'_n$ as the control cost of policy $\hat{u}_t = \theta_t(x_t)$ in the following system: 
	%Consequently, system \eqref{eq:LQG} controlled by the policy $u_t = \theta_t(x_t) + B^{\dagger} s_t$ is equivalent to the following system controlled by the policy $\hat{u}_t = \theta_t(x_t)$:
	\begin{align} \label{eq:trans-vec1}
		x_{t+1} = A x_t + B\hat{u}_t + w'_t,
	\end{align}
	where the noise $w'_t = Bs_t+w_t\sim \mathcal{N}(0,B\Phi_tB^\top + \Psi_w)$ is independent of $x_t$. Denote by $J_n^-$ the optimal control cost for system \eqref{eq:trans-vec1}. According to the certainty equivalence principle, the optimal feedback gain depends only on the system dynamics matrices $A$ and $B$, and is independent of the noise variance. Hence $\hat{u}_t = -K^{(n)}_t x_t$ is the optimal control policy for system \eqref{eq:trans-vec1}, where $K^{(n)}_t$ is the same as that defined in \eqref{eq:optKn}. It follows that
	\begin{align*}
		J^-_n  = \frac{1}{n}\sum_{t=1}^{n+1}\mathbb{E} \left[x^\top_tFx_t  \right] + \frac{1}{n}\sum_{t=1}^{n}\mathbb{E} \left[ (K^{(n)}_t x_t)^\top G  K^{(n)}_t x_t \right]   \le J'_n. 
	\end{align*}
	Now, applying the linear Gaussian policy $u_t = -K^{(n)}_t x_t + s_t$ to the original system \eqref{eq:LQG} yields the control cost 
	\begin{align*}
		J''_n &= \frac{1}{n}\sum_{t=1}^{n+1}\mathbb{E} [x^\top_tFx_t ]  +  \frac{1}{n}\sum_{t=1}^{n} \mathbb{E}  [u^\top_tGu_t ] \notag  \\
		& =  \frac{1}{n}\sum_{t=1}^{n+1}\mathbb{E} \left[x^\top_tFx_t   \right] + \frac{1}{n}\sum_{t=1}^{n}\mathbb{E} \left[ (K^{(n)}_t x_t )^\top G  K^{(n)}_t x_t \right] + \frac{1}{n}\sum_{t=1}^{n} \tr(G \Phi_t ) \\
		& = J^-_n + \frac{1}{n}\sum_{t=1}^{n} \tr(G \Phi_t) \\
		& \le J'_n + \frac{1}{n}\sum_{t=1}^{n} \tr(G \Phi_t)  = J_n.
	\end{align*}
	In addition, it is easy to see that
	\begin{align*}
		h(B(-K^{(n)}_t x_t + s_t) + w_t|x_t) = h(B(\theta_t(x_t) + s_t)+w_t|x_t) = h(Bs_t + w_t).
	\end{align*}
	Therefore, as far as the optimization problem \eqref{eq:cv-obj-ve}-\eqref{eq:cv-st-ve} is concerned,  both the policy $u_t = \theta_t(x_t) + s_t$ and the linear Gaussian policy $u_t = -K^{(n)}_t x_t + s_t$ achieve the same objective value. However, the linear Gaussian policy yields a smaller or equivalent constraint value. 
	This implies that $\bar{C}_n(V)$ defined in \eqref{eq:cv-obj-ve}-\eqref{eq:cv-st-ve} can be achieved by a linear Gaussian policy of the form ${u}_t = -K^{(n)}_t x_t + s_t$, where $s_t\sim \mathcal{N}(0,\Phi_t)$ is independent of $x_t$. The desired result is established.
	
\end{IEEEproof}

\subsection{Proof of Theorem \ref{thm:lowerbound}}  \label{subsec:proof-thm3}
This section presents the proof of Theorem \ref{thm:lowerbound}. 

\begin{IEEEproof}[Proof of Theorem \ref{thm:lowerbound}]
In general, by Fano's inequality, the capacity of the implicit channel admits the following expression:
\begin{align}
    C_{\text{no}}(V) = \lim_{n\to \infty} \ \max_{u_t(o^t,m): J_n\le J^*_n + V} \ \frac{1}{n} \mathcal{I}(m;z^{n+1}).
\end{align}
The lower bound is established by restricting our attention to the set of stationary linear Gaussian policies. Specifically, we consider input policies with a time-invariant feedback gain of the form\footnote{In fact, it suffices to consider policies with a time-invariant feedback gain and time-varying covariance for $s_t$. It can be shown that if the feedback gain is time-invariant, say $\bar{K}$, then the optimal covariance of $s_t$ corresponding to $\bar{K}$ is also time-invariant. For simplicity of the proof, we skip this step and assume that both $K_t$ and $\Phi_t$ are time-invariant.}  $u_t = -\bar{K}\check{x}_{t|t}+ s_t$, where $s_t\sim \mathcal{N}(0,\Phi)$ is independent of $\check{x}_{t|t}$. Note that $u_t$ is determined by the controller's observation sequence $o^t$ and the message $m$. Since $s_t$ is independent of $\check{x}_{t|t}$, it must also be independent of $o^t$. Consequently, under this policy, the sequence $s^n$ is fully determined by message $m$. We thus have
    \begin{align*}
        \mathcal{I}(m;z^{n+1}) &= h(z^{n+1}) - h(z^{n+1}|m) \\
        & = h(z^{n+1}) - h(z^{n+1}|m,s^n) \\
        & = \sum_{t=1}^n [h(z_{t+1}|z^t) - h(z_{t+1}|z^t, s^n)] \\
        & \overset{(a)}{=} \frac{1}{2}\sum_{t=1}^n \left[\log \det \left(\cov(z_{t+1} - \hat{z}_{t+1}) \right) - \log \det \left(\cov(z_{t+1} - \bar{z}_{t+1}) \right) \right], 
    \end{align*}
    where $\hat{z}_{t+1} \triangleq \mathbb{E}[z_{t+1}|z^t]$ and $\bar{z}_{t+1} \triangleq \mathbb{E}[z_{t+1}|z^t, s^n]=\mathbb{E}[z_{t+1}|z^t, s^t]$. 
    Equality (a) holds because, under the linear Gaussian policy, both $\cov(z_{t+1} - \hat{z}_{t+1})$ and $\cov(z_{t+1} - \bar{z}_{t+1})$ are Gaussian. 

We next proceed to derive $\cov(z_{t+1} - \hat{z}_{t+1})$. Let $\hat{x}_{t|k}=\mathbb{E}[x_t|z^k]$ be the estimate of $x_t$ based on $z^k$. Then
\begin{align*}
    z_{t+1}-\hat{z}_{t+1} = D_r(x_{t+1} - \hat{x}_{t+1|t}) + v_{t+1}.
\end{align*}
Therefore, the key issue is to compute $\cov(x_{t+1} - \hat{x}_{t+1|t})$. As discussed around \eqref{eq:correlation}, since the controller's state estimation introduces an estimation error $e_t$ that is correlated with the system state $x_t$, the receiver cannot directly estimate $x_t$ from $z^t$ using the standard Kalman filter. Therefore, we instead consider the equivalent system \eqref{eq:system-rho-1}-\eqref{eq:system-rho-2}. 

In this new system, estimates of state $\hat{\rho}_{t|t} = \mathbb{E}[\rho_t|z^t]$ can be computed using the standard Kalman filter.
	 In particular, in the prediction stage, the Kalman filter computes $\hat{\rho}_{t+1|t} = \bar{A}_\rho \hat{\rho}_{t|t}$ and the associated estimation error covariance matrix
	\begin{align*}
		\Sigma_{t+1|t} \triangleq \cov(\rho_{t+1}-\hat{\rho}_{t+1|t})= \bar{A}_\rho \Sigma_{t|t} \bar{A}_\rho^\top + \bar{\Phi}+  \Psi_{\bar{w}} +  \Psi_{\bar{q}}.
	\end{align*}
    Note that, unlike the controller, the receiver does not know the input $u_t$, hence the estimation error covariance is related to $\bar{K}$ and $\bar{\Phi}$.
	Upon receiving the new observation $z_{t+1}$, the Kalman filter updates the estimate as follows: 
	\begin{align*}
		\hat{\rho}_{t+1|t+1} = \hat{\rho}_{t+1|t} + L_{t+1}(z_{t+1} - D_\rho \hat{\rho}_{t+1|t}),
	\end{align*}
	where $$L_{t+1} = \Sigma_{t+1|t} D_\rho^\top \left(D_\rho\Sigma_{t+1|t}D_\rho^\top  + \Psi_v \right)^{-1}.$$
	The associated estimation error is given by $$\Sigma_{t+1|t+1}= (I-L_{t+1}D_\rho)\Sigma_{t+1|t}.$$
	As $t\to\infty$, $\Sigma_{t+1|t}$ converges to $\Sigma_\rho$, which is determined by the DARE:
	\begin{align} \label{eq: Sigma-inf-rho}
		\Sigma_\rho = \bar{A}_\rho \left(\Sigma_\rho - \Sigma_\rho D_\rho^\top(D_\rho\Sigma_\rho D_\rho^\top + \Psi_v)^{-1}D_\rho\Sigma_\rho  \right)\bar{A}_\rho^\top + \bar{\Phi} + \Psi_{\bar{w}} +  \Psi_{\bar{q}}.
	\end{align}
	Accordingly, 
	\begin{align*}
		&L_\rho\triangleq \lim_{t\to \infty} L_t = \Sigma_\rho D_\rho^\top(D_\rho \Sigma_\rho D_\rho^\top + \Psi_v)^{-1}, \\
		&\tilde{\Sigma}_\rho \triangleq \lim_{t\to\infty}\Sigma_{t|t} = (I-L_\rho D_\rho)\Sigma_\rho.
	\end{align*}
Since $z_{t+1}-\hat{z}_{t+1} = D_\rho(\rho_{t+1} - \hat{\rho}_{t+1|t}) + v_{t+1}$,
it follows that
\begin{align}
    \cov(z_{t+1} - \hat{z}_{t+1}) = D_\rho \Sigma_{t+1|t}D_\rho^\top + \Psi_v.
\end{align}
Using the same argument yields
\begin{align}
    \cov(z_{t+1} - \bar{z}_{t+1}) = D_\rho \Pi_{t+1|t}D_\rho^\top + \Psi_v,
\end{align}
where $\Pi_{t+1|t}$ converges to $\Pi$, which is determined by
\begin{align*}
    \Pi = {A}_\rho \left(\Pi - \Pi D_\rho^\top(D_\rho\Pi D_\rho^\top + \Psi_v)^{-1}D_\rho\Pi  \right){A}_\rho^\top + \Psi_{\bar{w}} +  \Psi_{\bar{q}}.
	\end{align*}
Note that $\Pi$ is independent of $\Phi$.
Combining the above arguments yields
\begin{align*}
    \lim_{n\to \infty} \frac{1}{n} \mathcal{I}(m;z^{n+1}) = \frac{1}{2} \log \det(D_\rho \Sigma_\rho D_\rho^\top + \Psi_v) - \frac{1}{2}\log \det(D_\rho \Pi D_\rho^\top + \Psi_v).
\end{align*}

It is now clear that
\begin{align*}
    C_{\text{no}}(V) &= \lim_{n\to \infty} \ \max_{u_t(o^t,m): J_n\le J^{**}+V}  \frac{1}{n} \mathcal{I}(m;z^{n+1}) \\
    &\ge \lim_{n\to \infty} \ \max_{\bar{K},\Phi: J_n\le J^{**}+V}   \frac{1}{n} \mathcal{I}(m;z^{n+1}) \\
    & = \max_{\bar{K},\Phi: J_n\le J^{**}+V} \frac{1}{2} \log \det(D_\rho \Sigma_\rho D_\rho^\top + \Psi_v) - \frac{1}{2}\log \det(D_\rho \Pi D_\rho^\top + \Psi_v).
\end{align*}
    We can convert the control constraint using Lemma \ref{lem:JforK} and obtain the optimization presented in the theorem. This completes the proof.
\end{IEEEproof}

\subsection{Proof of Theorem \ref{thm: channel-trans}}  \label{subsec:proof-thm4}
This section presents the proofs of Theorem \ref{thm: channel-trans}. Before that, we state a useful lemma.

\begin{lemma} \label{lem: haty-rho}
		 Let $y_t = \rho_{t+1}-A_\rho \rho_t$ and $\hat{y}_t = \mathbb{E}[y_t|z^{t+1}]$. Then for any $t\ge 1$,
		\begin{align}  \label{eq:channel-yhat-rho}
			\hat{y}_t =(I-A_\rho Q_\rho)L_\rho[D_\rho(\bar{s}_t + \bar{w}_t + \bar{q}_{t+1} + A_\rho \tau_t) + v_{t+1}  ].
		\end{align}
\end{lemma}
\begin{IEEEproof}
    See Appendix B.
\end{IEEEproof}

The channel defined in Lemma \ref{lem: haty-rho} can be further simplified by showing that $(I-A_\rho Q_\rho)$ is invertible and $L_\rho$ is full rank. On this basis, Theorem \ref{thm: channel-trans} is established in two steps: (1) showing that the resulting output sequence retains the same information about $s^n$ as $z^{n+1}$; and (2) proving that the capacity of the channel is given by $f(\bar{K},V)$.

\begin{IEEEproof}[Proof of Theorem \ref{thm: channel-trans}] By definition, 
    \begin{align}  \label{eq:haty-rho2}
		\hat{y}_t = \mathbb{E}[{y}_t|z^{t+1}] & = \hat{\rho}_{t+1|t+1} - A_\rho \hat{\rho}_{t|t+1} \notag \\
		& = (I-A_\rho Q_\rho)L_\rho (z_{t+1}-D_\rho A_\rho \hat{\rho}_{t|t}).
	\end{align}
Lemma \ref{lem: haty-rho} shows that, under the policy $u_t = -K \check{x}_{t|t} + s_t$, the implicit channel can be translated into a Gaussian MIMO channel, as defined in \eqref{eq:channel-yhat-rho}. This channel can be further simplified by noting that: (i) $L_\rho \in \mathbb{R}^{2d_1 \times d_1}$ has rank $d_1$ (i.e., full rank); and (ii) $(I-A_\rho Q_\rho)$ is invertible. The second claim is easy to verify:
\begin{align}  \label{eq: IAQ}
	   I-{A_\rho}Q_\rho &= I-{A_\rho}(I-L_\rho D_\rho)\Sigma_\rho A_\rho^\top\Sigma_\rho^{-1} \notag \\
	& = I - (\Sigma_\rho -  \bar{\Phi} - \Psi_{\bar{w}} - \Psi_{\bar{q}}) \Sigma_\rho^{-1}  = (\bar{\Phi} + \Psi_{\bar{w}} +  \Psi_{\bar{q}}) \Sigma_\rho^{-1} \succ 0.
    \end{align}  

We proceed to show that $L_\rho$ is full rank.
Since $\rho_t=[x_t, e_t]$, we can also decompose $\hat{\rho}_{t+1|t}$ as follows:
$$\hat{\rho}_{t+1|t}=\mathbb{E}[\rho_{t+1}|z^{t}]=[\mathbb{E}[x_{t+1}|z^{t}], \mathbb{E}[e_{t+1}|z^{t}]]=[\hat{x}_{t+1|t}, \hat{e}_{t+1|t}].$$
Accordingly, the estimation error covariance can be written as
	\begin{align*}
		\Sigma_{t+1|t} = \cov(\rho_{t} - \hat{\rho}_{t+1|t} )= \begin{bmatrix}
			\cov(x_t-\hat{x}_{t+1|t}) & \mathbb{E}[(x_t - \hat{x}_{t+1|t})(e_t-\hat{e}_{t+1|t})^\top] \\
			\mathbb{E}[(e_t-\hat{e}_{t+1|t})(x_t - \hat{x}_{t+1|t})^\top] & \cov(e_t - \hat{e}_{t+1|t})
		\end{bmatrix} .
	\end{align*}
	We thus can express $\Sigma_\rho$ as a block matrix as follows:
	\begin{align*}
		\Sigma_\rho \triangleq \begin{bmatrix}
			\Sigma_{11} & \Sigma_{12} \\
			\Sigma_{21} & \Sigma_{22}
		\end{bmatrix},
	\end{align*}
	where $\Sigma_{11} = \cov(x_t-\hat{x}_{t+1|t}) \in \mathbb{R}^{d_1\times d_1}$ and $\Sigma_{22} = \cov(e_t - \hat{e}_{t+1|t})  \in \mathbb{R}^{d_1\times d_1}$ are both positive definite.
	As a result, by the definitions of $L_\rho$ and $\tilde{\Sigma}_\rho$ in the proof of Theorem \ref{thm:lowerbound}, we can also write
	\begin{align}  \label{eq:L-rho}
		L_\rho= \begin{bmatrix}
			L_{\rho1} \\
			L_{\rho2}
		\end{bmatrix} \triangleq  \begin{bmatrix}
			\Sigma_{11} D_r^\top (D_r\Sigma_{11} D_r^\top + \Psi_v)^{-1} \\
			\Sigma_{21} D_r^\top (D_r\Sigma_{11} D_r^\top + \Psi_v)^{-1} 
		\end{bmatrix}, 
	\end{align}
	and
	\begin{align} \label{eq: tildSigma-inf-rho}
		\tilde{\Sigma}_\rho = \begin{bmatrix}
			(I - L_{\rho1}D_r )\Sigma_{11} & (I - L_{\rho1}D_r) \Sigma_{12} \\
			(I - L_{\rho2}D_r) \Sigma_{21} & (I - L_{\rho2}D_r) \Sigma_{22} 
		\end{bmatrix}.
	\end{align}
	Since $D_r$ is invertible, both the two sub-matrices $L_{\rho1}$ and $L_{\rho2}$ are invertible, implying that $L_\rho$ is full rank.

    Based on the above arguments, the channel in  \eqref{eq:channel-yhat-rho} is equivalent to the following:
	\begin{align} \label{eq:channel-bary-rho}
		\bar{y}_t \triangleq L_\rho^\dagger (I-A_\rho Q_\rho)^{-1}\hat{y}_t & = D_\rho(\bar{s}_t + \bar{w}_t + \bar{q}_{t+1} + A_\rho \tau_t) + v_{t+1}, \notag \\
		& = D_r(Bs_t + w_t) + D_\rho A_\rho \tau_t + v_{t+1},
	\end{align}
	where $L_\rho^\dagger = (L_\rho^\top L_\rho)^{-1}L_\rho^\top$ is the left inverse of $L_\rho$. 
	
	Next, we proceed to show that this translation preserves all information about $s^n$. That is, $\mathcal{I}(s^n;\bar{y}^n) = \mathcal{I}(s^n;z^{n+1})$. 
	To see this, applying the chain rule of mutual information yields
    \begin{align*}
    	\mathcal{I}(s^n;z^{n+1},\bar{y}^n) = \mathcal{I}(s^n;\bar{y}^n) + \mathcal{I}(s^n;z^{n+1}|\bar{y}^n) = \mathcal{I}(s^n;z^{n+1}) + \mathcal{I}(s^n;\bar{y}^n|z^{n+1}).
    \end{align*}
    By definition, $\bar{y}^n$ is a function of $z^{n+1}$, we thus have $ \mathcal{I}(s^n;\bar{y}^n|z^{n+1})=0$. 
    Furthermore, since $(I - {A_\rho}Q_\rho)$ is invertible and $L_\rho$ is full rank, \eqref{eq:haty-rho2} implies that $z_{t+1}$ can be uniquely determined by $\bar{y}_t$ and $z^t$---as $\hat{\rho}_{t|t}$ is a function of $z^t$. Using this fact and the chain rule yields
    \begin{align}  \label{eq:no-info-loss}
    	\mathcal{I}(s^n;z^{n+1}|\bar{y}^n) = \mathcal{I}(s^n;z_1|\bar{y}^n) + \sum_{t=1}^{n} \mathcal{I}(s^n; z_{t+1}|z^t,\bar{y}^n)  = 0.
    \end{align}
    We thus have
    \begin{align*}
    	\mathcal{I}(s^n;z^{n+1},\bar{y}^n) = \mathcal{I}(s^n;\bar{y}^n) = \mathcal{I}(s^n;z^{n+1}).
    \end{align*}
    The above identity implies that, as a function of $z^{n+1}$, $\bar{y}^n$ retains all information about $s^n$.

	In the remaining of this proof, we derive the capacity of the channel in \eqref{eq:channel-bary-rho}.
	First, let $\beta_t = D_rw_t +  D_\rho A_\rho \tau_t + v_{t+1}$, then according to \eqref{eq:channel-bary-rho} and \eqref{eq:no-info-loss},
	\begin{align*}
		\frac{1}{n}\mathcal{I}(\bar{s}^n;\bar{y}^n) =\frac{1}{n}\mathcal{I}({s}^n;\bar{y}^n) 
        &= \frac{1}{n}\mathcal{I}({s}^n;\bar{y}^n,z^{n+1}) \\
        &=  \frac{1}{n}\sum_{t=1}^{n} \left[ h(\bar{y}_t|z^{t}) - h({\beta}_t|z^{t},{s}^n) \right]  \\
		& = \frac{1}{2n} \sum_{t=1}^{n} [\log \det ({\cov}(\bar{y}_t-{\tilde{y}}_t)) - \log \det ({\cov}({\beta}_t-{\tilde{\beta}}_t))], 
	\end{align*}
	where $\tilde{y}_t  \triangleq \mathbb{E}[\bar{y}_t|z^{t}]$ and $\tilde{\beta}_t  \triangleq \mathbb{E}[{\beta}_t|z^t,{s}^n] = \mathbb{E}[{\beta}_t|z^t,\bar{s}^n] $. Without loss of optimality, in the remainder of this proof, we assume all the Kalman filters are already in the steady-state; hence each estimation error covariance has converged to its steady-state value. Since $\hat{\rho}_{t|t}=\mathbb{E}[\rho_t|z^t]$ is a function of $z^t$, we have
	\begin{align*}
		\tilde{y}_t =\mathbb{E}[\bar{y}_t|z^{t}] &= \mathbb{E}[ D_\rho( \bar{s}_t +\bar{w}_t + \bar{q}_{t+1} + A_\rho \tau_t) + v_{t+1} | z^{t}] \\
		& = D_\rho A_\rho \mathbb{E}[( \rho_t - \hat{\rho}_{t|t}) |z^{t}] = 0.
	\end{align*}
	It follows that
	\begin{align*}
		{\cov}(\bar{y}_t-{\tilde{y}}_t) = \cov(\bar{y}_t) & =  [D_\rho (\bar{\Phi}  + \Psi_{\bar{w}} +  \Psi_{\bar{q}} + A_\rho \tilde{\Sigma}_\rho A_\rho^\top)D_\rho^\top + \Psi_v]  \\
		& =  D_\rho \Sigma_\rho D_\rho^\top + \Psi_v.
	\end{align*}
	Similarly,
	\begin{align*}
		\tilde{\beta}_t = \mathbb{E}[{\beta}_t|z^t,{s}^n] =D_\rho A_\rho (\mathbb{E}[ \rho_t | z^t,\bar{s}^n] - \hat{\rho}_{t|t} ).
	\end{align*}
	Let $\bar{\rho}_{t|t} = \mathbb{E}[ \rho_t | z^t,\bar{s}^n]=\mathbb{E}[ \rho_t | z^t,\bar{s}^{t-1}]$ and $\bar{\sigma}_t = \rho_t - \bar{\rho}_{t|t}$. Then
	\begin{align*}
		{\cov}({\beta}_t-{\tilde{\beta}}_t) & = \cov(D_\rho( \bar{w}_t + \bar{q}_{t+1} + A_\rho (\rho_t - \hat{\rho}_{t|t} - \bar{\rho}_{t|t} + \hat{\rho}_{t|t})) + v_{t+1}  ) \\
		& = \cov( D_\rho( \bar{w}_t + \bar{q}_{t+1} + A_\rho \bar{\sigma}_t) + v_{t+1}  ) \\
		& = D_\rho \Pi D_\rho^\top + \Psi_v,
	\end{align*}
	where $\Pi$ is the steady-state value of the estimation error covariance $\cov(\rho_t - \mathbb{E}[ \rho_t | z^{t-1},\bar{s}^{t-1}])$, which  is determined by the following DARE:
	\begin{align*}
		\Pi = {A}_\rho \left(\Pi - \Pi D_\rho^\top(D_\rho\Pi D_\rho^\top + \Psi_v)^{-1}D_\rho\Pi  \right){A}_\rho^\top + \Psi_{\bar{w}} +  \Psi_{\bar{q}}.
	\end{align*}
	Since Lemma \ref{lem:JforK} has established that the control constraint  is equivalent to an input power constraint,
	it is well-know that the capacity of the Gaussian channel $\bar{y}_t = D_rBs_t + \beta_t$ is given by
	\begin{align*}
		\lim_{n\to \infty} \max_{p(s^n)}  \frac{1}{n} I(s^n; \bar{y}^n) = \max_{\Phi\succeq 0} \frac{1}{2}\log \frac{\det (D_\rho \Sigma_\rho D_\rho^\top + \Psi_v)}{\det(D_\rho \Psi_\rho D_\rho^\top + \Psi_v)},
	\end{align*}
	subject to the input power constraint $\tr((B^\top \Gamma B+ {G}) \Phi) \le V - \tr((\bar{\Gamma}-\Gamma) (\Sigma_c - \tilde{\Sigma}_c))$. The resulting optimization problem is equivalent to $f(\bar{K},V)$, as established in Proposition \ref{prop:f(K,V)}. This completes the proof.

\end{IEEEproof}

\section{Conclusion and Future Work} \label{sec:con}

In this paper, we studied implicit communication in LQG control systems. By defining the control system as an implicit communication channel, we demonstrated that information can be transmitted from the controller to a receiver that observes the system state—without using explicit communication channels—while simultaneously maintaining the control cost within a given level. Specifically, information is encoded into the control inputs, and then decoded by the receiver from noiseless or noisy observations of the states. We formulated the implicit communication problem as a co-design of control and channel coding, and formalized the trade-off between control and communication as the implicit channel capacity, subject to a constraint on control performance. The main contributions of this paper are theoretical analysis of this implicit channel capacity in two settings. We started from the simple setting where both the controller and the receiver have noiseless observations of the system state, and derived a closed-form expression for the channel capacity. We then extended the analysis to the general setting where both the controller and the receiver have noisy observations, and established a lower bound on the channel capacity.

Our analysis reveals two key insights into implicit communication in LQG control systems. First, when the controller and receiver have noiseless observations, the capacity-achieving input policy satisfies a separation principle, meaning that the control and channel coding tasks can be addressed independently without loss of optimality. Second, under the linear Gaussian input policies, the implicit channel becomes equivalent to a Gaussian MIMO channel, allowing existing channel codes to be applied in implicit communication. These two properties significantly simplify the practical coding problem and demonstrate that implicit communication can be effectively realized in practical scenarios.

This work opens a promising new avenue of research at the intersection of control and communication, with many interesting problems remaining to be explored in future work. On the theoretical side, the current point-to-point implicit communication model can be extended to networked settings \cite{el2011network}. For instance, when there are multiple receivers, the implicit channel may take the form of a broadcast channel; when multiple controllers are involved—as in decentralized control systems \cite{DecLQG}—the channel becomes a multiple-access channel. These models are particularly relevant for multi-agent systems. Characterizing the capacities of such implicit channels is both interesting and fundamental from an information-theoretic perspective. 

On the application side, exploring how this implicit communication framework can be practically deployed in real-world scenarios—such as autonomous driving, swarm robotics, and human-robot interaction—is another promising direction. Moreover, its application in multi-agent systems raises a profound question: if one agent communicates implicitly with others at the cost of degrading its own control performance, can this communication, in turn, improve the performance of the overall system? Examples from nature suggest a positive answer, but a theoretical understanding of this phenomenon in artificial systems remains an open and challenging problem requiring further investigation.

\section*{Appendix A}
In this appendix, we present proofs for Section \ref{sec:perfect}.
\subsection*{A.1 Proof of Lemma \ref{lem:hatGamma}}
\begin{IEEEproof}
	Let $K$ be the optimal control gain and $\Gamma$ be the solution to the DARE, as defined in \eqref{eq:optK} and \eqref{eq:Gamma}, respectively. Then the optimal control cost, achieved by $u_t = -Kx_t$, is given by
	\begin{align*}
		J^* = \tr(\Gamma  \Psi_w) = \tr(\Lambda U^\top {\Gamma}U).
	\end{align*}
	Consider a sub-optimal control policy $u_t = -Kx_t + s_t$, where $s_t\sim \mathcal{N}(0,\Psi_s)$ is independent of $x_t$ and $w_t$. The control cost under this policy is given by (see discussions around \eqref{eq:bellman2} and \eqref{eq:cost-K-s} in the proof of Lemma \ref{lem:constraint})
	\begin{align} 
		J= J^* + \tr((B^\top \Gamma B+{G}) \Psi_s) = J^* + \tr(\hat{\Gamma}\hat{\Psi}_s) = J^* + \sum_{i=1}^{d_1}\hat{\Gamma}(i,i)\hat{\Psi}_s(i,i),
	\end{align}
	where $\hat{\Psi}_s = U^\top {\Psi}_s U$.
	Suppose that $\hat{\Gamma}$ has $k$ negative diagonal entries. Without loss of generality, assume that the first $k$ diagonal entries are negative. Consider a choice of  $\hat{\Psi}_s $ satisfying  $\hat{\Psi}_s (i,i)>0$ for $1\le i \le k$ and  $\hat{\Psi}_s (i,i)=0$ for $i>k$. Then the policy $u_t = -Kx_t + s_t$ results in a control cost:
	\begin{align} 
		J=  J^* + \sum_{i=1}^{k}\hat{\Gamma}(i,i)\hat{\Psi}_s(i,i) < J^*.
	\end{align}
	which is strictly smaller than $J^*$.
	However, this contradicts the assumption that $u_t=-Kx_t$ is an optimal policy while $u_t = -Kx_t + s_t$ is a sub-optimal policy. Therefore, the assumption that $\hat{\Gamma}$ has negative diagonal entries must be false, completing the proof.
	
\end{IEEEproof}

\subsection*{A.2 Proof of Lemma \ref{lem:constraint}}
\begin{IEEEproof}
	Let $s_t$ follow the Gaussian distribution $\mathcal{N}(0,\Phi)$. The control cost of the policy $u_t=-Kx_t + s_t$, denoted by $J$, can be determined by the Bellman equation:
	\begin{align*}
		f(x) + J = x^\top Fx +\mathbb{E} [u^\top G u] + \mathbb{E}[f(z)]  
	\end{align*}
	where $f(x)$ is the differential value function and $z$ is the next state. We can show that $f(x)=x\Gamma x^\top$, where $\Gamma$ is defined in \eqref{eq:Gamma}. To see this, substituting $f(x)=x\Gamma x^\top$ and $u=-Kx + s$ into the Bellman equation yields 
	\begin{align} \label{eq:bellman2}
		x^\top \Gamma x  + J  &= x^\top Fx  + x^\top K^\top G K x  +\mathbb{E}\left[ s_t^\top G   s_t \right] + \mathbb{E}[z^\top \Gamma z ] \notag \\
		& = x^\top Fx  + x^\top K^\top G K x  +\mathbb{E}\left[ s_t^\top G  s_t \right]  + x^\top(A-BK)^\top \Gamma  (A-BK)x + \mathbb{E}[(w+Bs)^\top \Gamma (w+Bs)] \notag \\
		& = x^\top [F+K^\top GK + (A-BK)^\top \Gamma (A-BK) ] x + \tr(G \Phi)  + \tr(\Gamma(\Psi_w + B\Phi B^\top)) .
	\end{align}
	Using the expression of $K$ defined in \eqref{eq:optK}, we can derive 
	\begin{align*}
		F+K^\top GK + (A-BK)^\top \Gamma (A-BK) = F+A^\top(\Gamma-\Gamma B(G+B^\top \Gamma B)^{-1}B^\top \Gamma)A = \Gamma.
	\end{align*}
	The last equality follows from the DARE given in \eqref{eq:Gamma}. This verifies that indeed $f(x)=x\Gamma x^\top$ and that
	\begin{align} \label{eq:cost-K-s}
		J=\tr(\Gamma(\Psi_w + B\Phi B^\top)) + \tr(G \Phi  ) = J^* + \tr((B^\top\Gamma B+{G}) \Phi).
	\end{align}
	This completes the proof.
\end{IEEEproof}

\subsection*{A.3 Proof of Theorem \ref{thm:cap-scalar}}
\begin{IEEEproof}
	According to \eqref{eq:cv2-obj-inf}-\eqref{eq:cv2-st-inf}, the capacity is given by
	\begin{align} 
		{C}(V) = \max_{\Phi\ge 0} \ & \frac{1}{2} \log \frac{\det \left(\Psi_w+ B{\Phi} B^\top\right)}{ \det(\Psi_w)}   \\  
		s.t. \ & \tr(\Phi (B^\top \Gamma B+ {G})) \le  V.
	\end{align}
	If the system is a scalar system, then $ \tr(\Phi (B^\top\Gamma B+{G})) = \Phi (B^2\Gamma+{G}) $ and 
    \begin{align*}
        \frac{\det \left(\Psi_w+ B{\Phi} B^\top\right)}{ \det(\Psi_w)} = 1 + \frac{B^2 \Phi}{\Psi_w}.
    \end{align*}
     By KKT conditions, the optimal value is achieved by a $\Phi$ that satisfies the equality of the constraint, that is $\Phi = V/(B^2\Gamma+{G})$. Substituting it into the objective function yields
	\begin{align*}
		C(V) = \frac{1}{2} \log \left(1+ \frac{B^2 V}{\Psi_w(B^2\Gamma+ {G})}\right) =  \frac{1}{2} \log \left(1+ \frac{V}{J^* + B^{-2}G\Psi_w}\right). 
	\end{align*}
	This completes the proof.
\end{IEEEproof}

\section*{Appendix B}
The appendix presents proofs for Section \ref{sec:no}.

\subsection*{B.1 Proof of Proposition \ref{prop:f(K,V)}}
\begin{IEEEproof}
    We need to show that, for any fixed $\bar{K}$, the optimization problem in Theorem \ref{thm:lowerbound} is equivalent to $f(\bar{K},V)$ defined in Proposition \ref{prop:f(K,V)}. Note that $\bar{\Gamma}$ is a constant once $\bar{K}$ is fixed. We restate the optimization problem in Theorem \ref{thm:lowerbound} below and refer it as $f_0(\bar{K},V)$:
    \begin{align} \label{eq:f0-obj}
		f_0(\bar{K},V) := \max_{\Phi} \ & \frac{1}{2}\log {\det (D_\rho \Sigma_\rho D_\rho^\top + \Psi_v)} - \frac{1}{2}\log {\det(D_\rho \Pi D_\rho^\top + \Psi_v)} \\ \label{eq:f0-st1}
		\text{s.t. } & \tr(({G}+B^\top \bar{\Gamma} B)\Phi )   + \tr((\bar{\Gamma} - \Gamma )(\Sigma_c - \tilde{\Sigma}_c)) \le V \\ \label{eq:f0-st2}
		& \Sigma_\rho = {A}_\rho \left(\Sigma_\rho - \Sigma_\rho D_\rho^\top(D_\rho\Sigma_\rho D_\rho^\top + \Psi_v)^{-1}D_\rho\Sigma_\rho  \right){A}_\rho^\top + \bar{\Phi} + \Psi_{\bar{w}} +  \Psi_{\bar{q}} \\
        & \Phi \succeq 0,
	\end{align}
    
	Since $f_0(\bar{K},V)$ and $f(\bar{K},V)$ share the same objective function, it suffices to show that they have the same optimal solution. First, define a function
	\begin{align*}
		g(\Sigma, \bar{\Phi}) \triangleq {A}_\rho \left(\Sigma - \Sigma D_\rho^\top(D_\rho\Sigma D_\rho^\top + \Psi_v)^{-1}D_\rho\Sigma  \right){A}_\rho^\top +  \bar{\Phi} + \Psi_{\bar{w}} +  \Psi_{\bar{q}}.
	\end{align*}
	Under this definition, the constraint \eqref{eq:f0-st2} becomes $g(\Sigma_\rho, \bar{\Phi}) = \Sigma_\rho$. We now relax this equality to an inequality:
	\begin{align} \label{eq:relax-dare}
		g(\Sigma_\rho, \bar{\Phi}) - \Sigma_\rho \succeq 0.
	\end{align}
	Since $\Psi_v\succ 0$, $D_\rho\Sigma_\rho D_\rho^\top + \Psi_v$ is PD if $\Sigma_\rho$ is PSD.
	Then by Schur complement for PSD matrices, \eqref{eq:relax-dare} is equivalent to 
	\begin{align*}
		D_\rho\Sigma_\rho D_\rho^\top + \Psi_v \succeq 0 \text{ and } 
		\begin{bmatrix}
			D_\rho\Sigma_\rho D_\rho^\top + \Psi_v & D_\rho\Sigma_\rho {A}_\rho^\top \\
			{A}_\rho\Sigma_\rho D_\rho^\top & {A}_\rho\Sigma_\rho {A}_\rho^\top + \bar{\Phi} + \Psi_{\bar{w}} +  \Psi_{\bar{q}} - \Sigma_\rho
		\end{bmatrix} \succeq 0.
	\end{align*}
	Therefore, problem $f(\bar{K},V)$ is equivalent to a relaxed version of $f_0(\bar{K},V)$, where the equality constraint \eqref{eq:f0-st2} is replaced by the relaxed inequality \eqref{eq:relax-dare}.
	
	Before proceeding, we state a standard result about the discrete-time algebraic Riccati equation. 
	
	\textbf{Fact 1:} Given that $g(\Sigma, \bar{\Phi}) - \Sigma = 0$ and $g(\Sigma', \bar{\Phi}') - \Sigma' = 0$. If $\bar{\Phi} \succeq \bar{\Phi}'$, then $\Sigma \succeq \Sigma'$. 
	
	The proof of Fact 1 is straightforward. Note that $\Sigma$ is the fixed point of the discrete recursion: $\Sigma_{t+1} = g(\Sigma_t, \bar{\Phi})$. If $\bar{\Phi} \succeq \bar{\Phi}'$ and initializing two recursions from the same PSD matrix $\Sigma_0=\Sigma'_0$, then we can show by induction that 
	\begin{align*}
		\Sigma_{t+1} = g(\Sigma_t, \bar{\Phi}) \succeq \Sigma'_{t+1} = g(\Sigma'_t, \bar{\Phi}'), \  \forall t\ge 0.
	\end{align*}
	It follows that $\Sigma = \Sigma_\infty \succeq \Sigma' = \Sigma'_\infty$.
	
	Now, suppose that a tuple $(\Sigma_0, \bar{\Phi}_1)$ is an optimal solution to $f(\bar{K},V)$. If $g(\Sigma_0, \bar{\Phi}_1) - \Sigma_0 = 0$, then this tuple is also an optimal solution to the problem $f_0(\bar{K},V)$. We consider the non-trivial case in which the inequality does not hold with equality:
	\begin{align*}
		\Delta \triangleq g(\Sigma_0, \bar{\Phi}_1) - \Sigma_0 \succeq 0.
	\end{align*}
	Let $\bar{\Phi}_0 = \bar{\Phi}_1 - \Delta$, then we have $g(\Sigma_0, \bar{\Phi}_0) - \Sigma_0 = 0$. For $\bar{\Phi}_1$, there also exists a PSD matrix $\Sigma_1$ satisfying $g(\Sigma_1, \bar{\Phi}_1) - \Sigma_1 = 0$. Since $\bar{\Phi}_1 \succeq \bar{\Phi}_0$, it follows from Fact 1 that $\Sigma_1 \succeq \Sigma_0$. Note that the tuple $(\Sigma_1, \bar{\Phi}_1) $ is a valid solution to both problems $f(\bar{K},V)$ and $f_0(\bar{K},V)$. It follows from $D_\rho\Sigma_1 D_\rho^\top + \Psi_v \succeq D_\rho\Sigma_0 D_\rho^\top + \Psi_v \succ 0$ that
	\begin{align*}
		\log {\det ( D_\rho\Sigma_1 D_\rho^\top + \Psi_v) }  \ge  \log {\det ( D_\rho\Sigma_0 D_\rho^\top + \Psi_v }).
	\end{align*}
	Since $(\Sigma_0, \bar{\Phi}_1)$ is assumed to be an optimal solution to $f(\bar{K},V)$, the above inequality must hold with equality.
	
	We thus obtain the following conclusion: if $(\Sigma,\bar{\Phi})$ is an optimal solution to problem $f(\bar{K},V)$, then one of the following must hold:
	\begin{itemize}
		\item [(i)] $g(\Sigma, \bar{\Phi}) - \Sigma = 0$, in which case $(\Sigma,\bar{\Phi})$ is also an optimal solution to problem $f_0(\bar{K},V)$.
		\item [(ii)] There exists a PSD matrix $\Sigma'\succeq \Sigma$ such that $g(\Sigma', \bar{\Phi}) - \Sigma' = 0$ and $(\Sigma',\bar{\Phi})$ yields the same objective value. In this case, $(\Sigma',\bar{\Phi})$ is an optimal solution to both problems.
	\end{itemize}
	In both cases, the two optimization problems share an identical optimal solution. 
    
    To verify that the problem $f(\bar{K},V)$ is a convex optimization \cite{boyd2004convex}, note that: (i) the second term of the objective is a constant and does not depend on the decision variables; (ii) the function $\log \det (\cdot)$ is concave; (iii) all constraints are linear.
    This completes the proof.
	
\end{IEEEproof}

\subsection*{B.2 Proof of Lemma \ref{lem:JforK}}
\begin{IEEEproof}
	For ease of notation, we will write $\check{x}_{t|t}$ as $\check{x}_t$ whenever the context is clear. We prove this lemma using a similar argument as in Lemma \ref{lem:constraint}. That is, we begin by assuming that the differential value function under the policy $u_t  = -\bar{K}\check{x}_{t} + s_t$ takes the quadratic form $f(\check{x}_t) = \check{x}_t^\top \bar{\Gamma} \check{x}_t$. Then we show that this function indeed satisfies the Bellman equation. The corresponding average control cost is subsequently derived from the Bellman equation. First, substituting $f(\check{x}_t) = \check{x}_t^\top  \bar{\Gamma}\check{x}_t$ into the Bellman equation yields
	\begin{align} \label{eq:bellman-check}
		\check{x}_t^\top \bar{\Gamma} \check{x}_t  + J  &= (\check{x}_t+e_t)^\top F(\check{x}_t +e_t)  + \check{x}_t^\top \bar{K}^\top G \bar{K} \check{x}_t  +\mathbb{E}\left[ s_t^\top G  s_t \right] + \mathbb{E}[\check{x}_{t+1}^\top \bar{\Gamma} \check{x}_{t+1} ] .
	\end{align}
	Without loss of optimality, we assume the Kalman filter is already in the steady-state. Then, according to the Kalman filter formulas,
	\begin{align*}
		\check{x}_{t+1} & = (I-L_cD_c)(A\check{x}_t + Bu_t )+ L_c o_{t+1} \\
		& = (I-L_cD_c)(A\check{x}_t + Bu_t )+ L_c (D_c(Ax_t + Bu_t +w_t) + q_{t+1})  \\
		& = A\check{x}_t + Bu_t + L_cD_c (Ae_t + w_t) + L_c q_{t+1}.
	\end{align*}
	Note that $Ae_t + w_t$ is uncorrelated with $A\check{x}_t + Bu_t$ and that $\cov(D_c (Ae_t + w_t))=D_c(A \tilde{\Sigma}_cA^\top +\Psi_w)D^\top_c = D_c\Sigma_c D^\top_c$. It follows that
	\begin{align*}
		\mathbb{E}[\check{x}_{t+1}^\top \bar{\Gamma} \check{x}_{t+1} ]
		&=\mathbb{E} [( A\check{x}_t + Bu_t )^\top \bar{\Gamma} ( A\check{x}_t + Bu_t )] +  \tr(\bar{\Gamma} L_c (D_c \Sigma_{c}D_c^\top + \Psi_q) L_c^\top) \\
		& = \check{x}^\top_t (A-B\bar{K})^\top \bar{\Gamma} (A-B\bar{K})\check{x}_t + \mathbb{E}[(Bs_t)^\top \bar{\Gamma} Bs_t]+  \tr(\bar{\Gamma} L_c (D_c \Sigma_{c}D_c^\top + \Psi_q) L_c^\top). 
	\end{align*}
	Since $L_c = \Sigma_cD_c^\top (D_c\Sigma_cD_c^\top +\Psi_q)^{-1}$ and $\tilde{\Sigma}_c = (I-L_c D_c)\Sigma_c$, we have
	$$\tr(\bar{\Gamma} L_c (D_c \Sigma_{c}D_c^\top + \Psi_q) L_c^\top) = \tr(\bar{\Gamma}\Sigma_cD_c^\top L_c^\top) = \tr(\bar{\Gamma} (\Sigma_c - \tilde{\Sigma}_c)). $$
	Substituting the above into \eqref{eq:bellman-check} yields
	\begin{align} \label{eq:bellman-check-2}
		\check{x}_t^\top \bar{\Gamma} \check{x}_t  + J  =&\check{x}^\top_t  [F+\bar{K}^\top G\bar{K} + (A-B\bar{K})^\top \bar{\Gamma} (A-B\bar{K}) ] \check{x}_t + \mathbb{E}[e_t^\top F \check{x}_t] + \mathbb{E}[\check{x}_t^\top F e_t] \notag \\
		&+ \tr(G \Phi) + \tr(F\tilde{\Sigma}_c) + \tr(B^\top \bar{\Gamma} B \Phi) + \tr(\bar{\Gamma} (\Sigma_c - \tilde{\Sigma}_c)).
	\end{align}
	It is well-known that, in Kalman filter, the estimation error $e_t$ and the estimate of state $\check{x}_t$ are uncorrelated. We thus have $\mathbb{E}[e_t^\top F \check{x}_t] = \mathbb{E}[\check{x}_t^\top F e_t]=0$.
	Furthermore, as discussed in the  proof of Lemma \ref{lem:constraint}, \eqref{eq:bellman-check-2} holds if
	$$ \bar{\Gamma} = F+\bar{K}^\top G\bar{K} + (A-B\bar{K})^\top \bar{\Gamma} (A-B\bar{K}).$$
	This verifies that $f(\check{x}_t) = \check{x}_t^\top \bar{\Gamma} \check{x}_t$ is the differential value function. It follows immediately from the Bellman equation \eqref{eq:bellman-check-2} and Eq. \eqref{eq:opt-J-noise} that  
	$$J = \tr(G \Phi) + \tr(F\tilde{\Sigma}_c) + \tr(B^\top \bar{\Gamma}B\Phi) + \tr(\bar{\Gamma} (\Sigma_c - \tilde{\Sigma}_c)) = J^{**} + \tr((B^\top \bar{\Gamma} B+ {G}) \Phi) + \tr((\bar{\Gamma} -\Gamma) (\Sigma_c - \tilde{\Sigma}_c)) . $$
	This completes the proof.
\end{IEEEproof}

\subsection*{B.3 Proof of Lemma \ref{lem: haty-rho}}
\begin{IEEEproof}
	First, it is easy to see that $\{\tau_t:t\ge 1\}$ forms a Markov chain:
	\begin{align}  
		\tau_{t+1} &= \rho_{t+1} - \hat{\rho}_{t+1|t+t}   \notag \\
		& = {A_\rho}\rho_t + \bar{s}_t + \bar{w}_t + \bar{q}_{t+1} - {A_\rho}\hat{\rho}_{t|t} - L_\rho (z_{t+1} - D_\rho{A_\rho}\hat{\rho}_{t|t} )  \notag  \\
		& = (I-L_\rho D_\rho){A_\rho} \tau_t - L_\rho v_{t+1} + ( I-L_\rho  D_\rho) (\bar{s}_t + \bar{w}_t + \bar{q}_{t+1}).
	\end{align}
	The last line holds because $D_\rho{A_\rho}{\rho}_{t} = D_\rho(\rho_{t+1} - \bar{s}_t - \bar{w}_t - \bar{q}_{t+1}) = z_{t+1} - v_{t+1} - D_\rho (\bar{s}_t + \bar{w}_t + \bar{q}_{t+1}) $. We note that $\tau_t$ depends on $z^t$ and is independent of $s_t,w_t$, $q_t$, and $v_{t+1}$.
	
	Let $\varepsilon_t \triangleq \rho_t -  \hat{\rho}_{t|t+1} $ denote the estimation error of the Kalman smoother. Then according to \eqref{eq:smoother-rho},
	\begin{align*}
		\varepsilon_t &=\rho_t -  \hat{\rho}_{t|t} - Q_\rho \left(\hat{\rho}_{t+1|t+1} - {A_\rho} \hat{\rho}_{t|t}\right)  \\
		& = \tau_t - Q_\rho L_\rho (z_{t+1} - D_\rho{A_\rho}\hat{\rho}_{t|t} )  \\
		& = (I - Q_\rho L_\rho D_\rho {A_\rho})\tau_t - Q_\rho L_\rho(v_{t+1} + D_\rho (\bar{s}_t + \bar{w}_t + \bar{q}_{t+1})).
	\end{align*}
	Define the estimation error $\eta_t \triangleq y_t - \hat{y}_t $. Then it can be written as
	\begin{align*}
		\eta_t & = \rho_{t+1} - \hat{\rho}_{t+1|t+1} - {A_\rho} \left(\rho_t  - \hat{\rho}_{t|t+1} \right)  \\
		& = \tau_{t+1} - {A_\rho} \varepsilon_t \\ 
		& = ({A_\rho}Q_\rho-I)L_\rho D_\rho{A_\rho} \tau_t + ({A_\rho}Q_\rho - I)L_\rho v_{t+1} + ( {A_\rho}Q_\rho L_\rho D_\rho + I -L_\rho D_\rho) (\bar{s}_t + \bar{w}_t + \bar{q}_{t+1}).
	\end{align*}
	We thus have
	\begin{align*}
		\hat{y}_t & = y_t - \eta_t = \bar{s}_t + \bar{w}_t + \bar{q}_{t+1} - \eta_t  \\
		& = (I-{A_\rho}Q_\rho)L_\rho D_\rho {A_\rho} \tau_t + (I-{A_\rho}Q_\rho)L_\rho v_{t+1} + (I - {A_\rho}Q_\rho)L_\rho D_\rho (\bar{s}_t + \bar{w}_t + \bar{q}_{t+1}) \\
		& = (I - {A_\rho}Q_\rho)L_\rho  [D_\rho(\bar{s}_t + \bar{w}_t + \bar{q}_{t+1} + A_\rho \tau_t) + v_{t+1}  ].
	\end{align*}
    This completes the proof.
\end{IEEEproof}

\bibliographystyle{IEEEtran}
\bibliography{reference}

\end{document}